\documentclass[12pt, a4paper]{article}
\pdfoutput=1
\usepackage{color, amsthm}
\usepackage[centertags,reqno]{amsmath}
\usepackage{amssymb,amscd}
\usepackage[utf8]{inputenc}
\usepackage{hyperref}
\usepackage{titlesec}
\usepackage{appendix}

\usepackage{microtype}

\DeclareMathOperator{\End}{End}
\DeclareMathOperator{\Id}{Id}
\DeclareMathOperator{\lie}{Lie}
\def\cO{\mathcal{O}}
\def\vac{|0\rangle}

\hypersetup{
 pdftitle={ Chiral de Rham complex on  Riemannian manifolds and special holonomy},    % title
    pdfauthor={Joel Ekstrand, Reimundo Heluani, Johan Kallen and Maxim Zabzine}, 
    pdfnewwindow=true,      % links in new window
    colorlinks=true,       % false: boxed links; true: colored links
    linkcolor=black,          % color of internal links
    citecolor= black,        % color of links to bibliography
    filecolor= black,      % color of file links
    urlcolor= black           % color of external links
}

\numberwithin{equation}{section} 

\newtheorem{theorem}{Theorem}[section]
\newtheorem{prop}[theorem]{Proposition}
\newtheorem{conj}[theorem]{Conjecture}
\newtheorem{ex}{Example}[section]
\newtheorem{lem}[theorem]{Lemma}
\theoremstyle{definition}
\newtheorem{defn}[theorem]{Definition}
\newtheorem{rem}[theorem]{Remark}

\def\+{{+\!\!\!+}} 
\def\d{\partial} 
\def\G{\Gamma}   

\def\L{\Lambda} 
\newcommand{\beq}{\begin{equation}} 
\newcommand{\eeq}[1]{\label{#1}\end{equation}}
\newcommand{\bs}{\begin{split}}
\newcommand{\es}{\end{split}}
\newcommand{\ie}{{i.e.\ }}

\newcommand{\uD}{D}
\newcommand{\ud}{d}

\newcommand{\LB}[3][\Lambda]{[\, #2  \,_{#1}\,  #3 \,]}
\newcommand{\PLB}[3][\Lambda]{\{\, #2  \,_{#1}\,  #3 \,\}}

\begin{document}

\def\cR{\mathcal{R}}
\newcommand{\inv}[1]{{#1}^{-1}} %inverse 
\newcommand{\be}{\begin{equation}} 
\newcommand{\ee}{\end{equation}} 
\newcommand{\bea}{\begin{eqnarray}} 
\newcommand{\eea}{\end{eqnarray}} 
\newcommand{\re}[1]{(\ref{#1})} 
\newcommand{\qv}{\quad ,} 
\newcommand{\qp}{\quad .}

\newcommand{\spins}{\mathrm{Spin}(7)}

\begin{flushright} \small
UUITP-08/10  
 \end{flushright}
\smallskip
\begin{center} \LARGE
{\bf  Chiral de Rham complex on\\
  Riemannian manifolds and special holonomy}
  \\[12mm] \normalsize
{\bf Joel~Ekstrand$^a$, Reimundo~Heluani$^b$, \\
Johan~K\"all\'en$^a$ and Maxim Zabzine$^a$} \\[8mm]
 {\small\it
 $^a$Department of Physics and Astronomy, 
     Uppsala university,\\
     Box 516, 
     SE-751\;20 Uppsala,
     Sweden\\
     ~\\
     $^b$Department of Mathematics, University of California,\\
      Berkeley, CA 94720, USA
}
\end{center}
\vspace{7mm}

\begin{abstract}
\noindent  Interpreting the chiral de Rham complex (CDR) as a formal 
Hamiltonian quantization of the supersymmetric non-linear sigma model, we suggest a setup for 
 the study of CDR on manifolds with special holonomy. 
 We show how to systematically construct global sections of CDR from differential forms, and investigate the algebra of the sections corresponding to the covariantly constant forms associated with the special holonomy.
 As a concrete example, we construct two commuting copies of the Odake algebra (an extension of the $N=2$  superconformal algebra)
  on the space of global sections of CDR of a Calabi-Yau threefold and conjecture similar results for $G_2$ manifolds. 
We also discuss quasi-classical limits of these algebras.
\end{abstract}

\section{Introduction}
The chiral de Rham complex (CDR) was introduced by Malikov, Schechtman and Vaintrob in \cite{Malikov:1998dw}. 
CDR is a sheaf of  supersymmetric vertex algebras over a smooth manifold $M$. It is defined on a local coordinate patch of $M$ as a $bc$--$\beta\gamma$-system --- a tensor product of several copies of the Clifford vertex algebra
 % (free Fermionic ghosts $bc$)
   and the Weyl vertex algebra,
  % (free Bosonic ghosts $\beta\gamma$ )
   and then extended to $M$ by gluing on the intersections of these coordinate patches.  Since the original work \cite{Malikov:1998dw}, there has been considerable 
   progress in understanding the mathematical aspects of CDR (cf. \cite{MR2038198, benzvi-2006, Malikov:2006rm} among others). 
    In the physics literature, CDR appeared in  the context of half-twisted sigma models
      \cite{Kapustin:2005pt, Witten:2005px} and in the context 
     of infinite volume limits of sigma models \cite{Frenkel:2005ku, Frenkel:2006fy, Frenkel:2008vz}.  The present work is the logical continuation of \cite{Ekstrand:2009zd},
      where it was suggested to interpret CDR as a formal canonical quantization of the non-linear sigma model. 

In the original work \cite{Malikov:1998dw}, the authors showed that if $M$ is Calabi-Yau, the vertex algebra of global sections of CDR admits an embedding of the $N=2$ superconformal vertex algebra. This result was further extended in \cite{benzvi-2006} showing that in the hyperk\"ahler case, one obtains the $N=4$ superconformal vertex algebra as a subalgebra of global sections of CDR. Later, it was shown in \cite{heluani-2008} that in fact in both cases one obtains two commuting copies of the corresponding supersymmetry algebras, each with half the central charge. These results parallel the corresponding statements in the physics literature, where one expects $N=2$ (resp. $N=4$) superconformal symmetry in the sigma model with target a Calabi-Yau (resp. hyperk\"ahler) manifold. Starting with the observation by P.~Howe and G.~Papadopoulos in  \cite{Howe:1991vs, Howe:1991ic}, 
    where the relation between classical symmetries of non-linear sigma models and 
     special holonomy manifolds  was observed, this relation has now become well established in the physics literature.
     Also, when $M$ is an affine space the algebras associated to  Calabi-Yau threefolds, and to  $G_2$ and $\spins$ manifolds, have been studied in a quantum treatment of the sigma model by Odake in \cite{Odake:1988bh} and by Shatashvili  and Vafa in \cite{Shatashvili:1994zw}, respectively.
     
%     $M$ is a Calabi-Yau threefold, we explicitly construct an embedding of the algebra studied by Odake \cite{Odake:1988bh}
     
	 It is natural to ask if these observations reflect in a relation between special holonomy of $M$ and the existence of certain subalgebras of CDR. In this article we provide some technical tools to answer this question. We describe an embedding (Theorem \ref{thm:3}), different than the one in \cite{Malikov:1998dw}, of the space of differential forms $\Omega^*(M)$ into global sections of CDR. When the manifold $M$ has special holonomy it admits covariantly constant forms, and we obtain their corresponding sections of CDR and the subalgebra generated by them. 
	 In the case when $M$ is a Calabi-Yau threefold, we explicitly construct in this way an embedding of the algebra studied by Odake \cite{Odake:1988bh} into the space of global sections of CDR. 
%	  In some cases, these sections generate a sub-vertex algebra of CDR which is isomorphic to the corresponding symmetry algebra predicted by physicists in \cite{Odake:1988bh,Shatashvili:1994zw} (see Section \ref{s:prelim} for their description in vertex algebra language). 
	 
	 The relation of these symmetries of CDR with those of the classical sigma model is more than just an analogy. There have been rigorous works in the mathematical literature explaining this, see  for example \cite{Malikov:2006rm}  where a \emph{quasi-classical limit} of CDR is considered. 
	%We study this relation in the spirit of \cite{Malikov:2006rm}.
	 In fact, CDR as a sheaf of vertex algebras is a quantization of a sheaf of Poisson vertex algebras (see Section \ref{s:prelim} for a definition). Roughly speaking, a Poisson vertex algebra is obtained as a limit of a family of vertex algebras $V_\hbar$, such that, in the limit $\hbar \rightarrow 0$, the $(-1)$-product becomes commutative.  The resulting limit $V_0 := \lim_{\hbar \rightarrow 0} V_{\hbar}$ inherits its commutative product from the $(-1)$-product of $V_\hbar$ and a Lie conformal structure from the OPE of $V_{\hbar}$. In this sense, we can put CDR in a family parametrized by $\hbar$ in such a way that its limit becomes a Poisson vertex algebra. 
	 We show in this article how all the above mentioned symmetries of the classical sigma model in the physics literature translate verbatim to symmetries of these Poisson vertex algebras. Moreover, the approach in this article shows in which sense the symmetries of CDR obtained in \cite{Malikov:1998dw,benzvi-2006,heluani-2008} and in our main Theorem \ref{main:theorem} are quantizations of the corresponding symmetries of the classical sigma model, since the sections of CDR corresponding to covariantly constant forms on $M$ become the generators of the classical symmetries when $\hbar = 0$.  

 The article is  organized as follows: in Section \ref{s:prelim} we set up notation and give the necessary background on (SUSY) vertex algebras and Poisson vertex algebras. We give here all the examples of vertex algebras and their corresponding Poisson versions considered in the rest of the article. 

The results of the next two sections \ref{s:classical} and \ref{clAlgExt} should be of interest for the more physically inclined reader, however these sections could be skipped on a first reading since the main results in the quantum setup of section \ref{lift} are independent of these sections. 
	 In Section \ref{s:classical} we review the Hamiltonian formalism for $N=(1,1)$ 
  supersymmetric non-linear sigma models. In this section we sketch a dictionary between Lambda brackets on Poisson vertex algebras and Poisson brackets of local functionals on loop spaces. 
  Section \ref{clAlgExt} deals with the classical symmetries of sigma models on  special holonomy manifolds.  
   This section presents the Hamiltonian treatment of the results from \cite{Howe:1991vs, Howe:1991ic}. Using the dictionary between local functionals and Poisson vertex algebras, we give a list of Poisson vertex algebras associated to the different cases of special holonomy. In Section \ref{s:CDR} we recall  the definition of 
      CDR.  In Section \ref{lift} we prove the main technical result of this article, Theorem \ref{thm:3}, which produces two non-trivial embeddings of the spaces of differential forms on a Riemannian manifold $M$ into global sections of CDR. This embedding depends explicitly on the Levi-Civita connection on $M$. It is shown here that taking a quasi-classical limit $\hbar \rightarrow 0$ we recover from these sections the classical currents considered in the previous section \ref{clAlgExt}, therefore CDR provides us with a natural setup to quantize the classical symmetries of the sigma model. 
 Section \ref{CDRalgExt} presents our results at the quantum level. The main result is the construction of
  two commuting copies of the Odake algebra on a  Calabi-Yau threefold.  We also present a conjecture in the $G_2$  case (conjecture \ref{conj1}), and discuss the $\spins$ case.
 In Section \ref{s:summary} we present a summary and discuss the main complications in further possible calculations. 
 In 
Appendix \ref{crosscontractions} we collect some  useful formulas on different special holonomy manifolds.  

\section{Algebraic preliminaries} \label{s:prelim}
In this section we collect the necessary tools from the theory of vertex algebras and SUSY vertex algebras. For details the reader is referred to \cite{Heluani:2006pk}. We give all the examples of vertex algebras and vertex Poisson algebras that we will need in the rest of the article. The SUSY vertex algebras in this article are called SUSY $N_K=1$ vertex algebras in \cite{Heluani:2006pk}. We will drop the $N_K=1$ moniker.  The reader might benefit also from \cite{benzvi-2006} where the formalism of \cite{Heluani:2006pk} is described in the particular case of $N_K=1$ SUSY vertex algebras. 

All modules and algebras in this article are $\mathbb{Z}/2\mathbb{Z}$-graded and we will often drop the word super when no confusion could arise. For an element $a$ in a super vector space of degree $\Delta_a \in \mathbb{Z}/2 \mathbb{Z}$, we will denote $(-1)^{\Delta_a}$ by  $(-1)^a$. 

Consider the $1|1$ dimensional super Lie algebra with an odd generator $D$ and an even generator $\partial$ with commutation relations
\def\cH{\mathcal{H}}
\begin{equation}
\label{eq:m1.1}
[D, D] = 2\partial, \qquad [\partial,D] = [\partial,\partial] = 0,
\end{equation}
and let $\cH$ be its universal enveloping algebra. We will also consider another set of generators $\chi, \lambda$ for $\cH$ such that $\chi^2 = - \lambda$. We will denote $\nabla=(\partial,D)$ and $\Lambda =  (\lambda, \chi)$. 
\begin{defn}
    \label{defn:k.conformal.1}
%\label{defn:1}
A SUSY Lie conformal algebra is a $\cH$-module $\cR$ together with a degree $1$ operation
\begin{equation}
\label{eq:m1.2}
[\cdot_\Lambda \cdot] : \cR \otimes \cR \rightarrow \cH \otimes \cR,
\end{equation}
called the Lambda bracket satisfying 
\begin{enumerate}
        \item Sesquilinearity
            \begin{equation}
                [D a_\Lambda b] =  \chi [a_\Lambda b]~,
                \qquad [a_\Lambda D b] = -(-1)^{a} \left(D
                + \chi
                \right) [a_\Lambda b]~.
                \label{eq:k.sesqui.1}
            \end{equation}
			Here we use the elements $(\partial, D)$ of $\mathcal{H}$ acting on $\mathcal{R}$ while $(\lambda, \chi)$ denote elements of the first factor in the right hand side of  \eqref{eq:m1.2}.
        \item Skew-symmetry:
            \begin{equation}
                [b_\Lambda a] =  (-1)^{a b} [b_{-\Lambda -
                \nabla} a]~.
                \label{eq:k.skew.1}
            \end{equation}
            Here the bracket on the right hand side is computed as
            follows: first compute $[b_{\Gamma}a]$, where $\Gamma =
            (\gamma, \eta)$ are generators of $\cH$. This way one obtains a polynomial in $(\gamma, \eta)$ with coefficients in $\mathcal{R}$. Next replace $\Gamma$ by $(-\lambda - \partial,
            -\chi - D)$, and as above, apply $(\partial, D)$ to $\mathcal{R}$. This way both sides of \eqref{eq:k.skew.1} are polynomials in $(\lambda, \chi)$ with coefficients in $\mathcal{R}$. A more detailed explanation along with commuting diagrams can be found in \cite[Rem. 4.11]{Heluani:2006pk}.
        \item Jacobi identity:
            \begin{equation}
                [a_\Lambda [b_\Gamma c]] = -(-1)^{a} \left[
                [ a_\Lambda b]_{\Gamma + \Lambda} c \right] +
                (-1)^{(a+1)(b+1)} [b_\Gamma [a_\Lambda c]]~,
                \label{eq:k.jacobi.1}
            \end{equation}
            where the first bracket on the right hand side is computed as in Skew-Symmetry
            and the identity is an identity in $\cH^{\otimes 2} \otimes\cR$.
    \end{enumerate}
\end{defn}

\begin{ex}[$N=1$ superconformal algebra] 
\label{ex:1}
This is the $\cH$ module $\mathcal{R}$ which is the direct sum of a free module generated by an odd vector $T$ and a one dimensional module spanned by an even element $C$ such that $DC = 0$ and the non-trivial bracket is given by:
\begin{equation}
\label{eq:m1.3}
\LB{T}{T} = (2 \partial + \chi D + 3 \lambda) T + \frac{\lambda^2 \chi}{3} C.
\end{equation}
The element $T$ is often called \emph{super Virasoro} or \emph{Neveu-Schwarz} field. Note that $C$ satisfies $[C_\Lambda \mathcal{R}] = 0$. We will call such an element central. 
\end{ex}
\noindent 
Let $T$ be as in \eqref{eq:m1.3}, 
following definition is \cite[Def. 5.6]{Heluani:2006pk} we say that  
a vector $a$ is said to have conformal weight $\Delta \in \mathbb{C}$ with respect to the $N=1$ generator $T$ if it satisfies:
\begin{equation}
\LB{T}{a}  = (2 \partial + \chi D + 2 \Delta \lambda) a + \mathcal{O}(\lambda \chi)~.
\label{eq:m1.10}
\end{equation}
Where $O(\lambda \chi)$ denote higher order terms in the PBW basis of $\mathcal{H}$ given by $(\lambda, \chi)$. 
It is said to be primary if these terms vanish.

\begin{ex}[$N=2$ superconformal algebra] 
\label{ex:2}
This is an extension of the $N=1$ algebra by an even vector $J$ satisfying
\begin{equation}
\begin{aligned}
\LB{T}{J} &= (2\partial+ \chi D + 2\lambda) J, \\
\LB{J}{J} &= - \left( T + \frac{C}{3} \lambda\chi \right).
\end{aligned}
\label{eq:m1.4}
\end{equation}
\end{ex}

\begin{ex}[$N=4$ superconformal algebra] 
\label{ex:3}
This algebra is generated by an odd vector $T$, three vectors $\{J^i\}_{i=1}^3$ and a central element $C$ such that each triple $(T, J^i, C)$ is a copy of the $N=2$ algebra and the $J^i$ satisfy:
\begin{equation}
\LB{J^i}{ J^j} = \varepsilon_{ijk} (\chi + 2 D) J^k,
\label{eq:m1.5}
\end{equation}
where $\varepsilon_{ijk}$ is the totally antisymmetric tensor. 
\end{ex}

\def\vac{|0\rangle}
\begin{defn}
\label{defn:}
A SUSY vertex algebra is a tuple $(V, \vac, \cdot, [\cdot_\Lambda \cdot])$ where $(V, [\cdot_\Lambda \cdot])$ is a SUSY Lie conformal algebra, $\vac \in V$ is an even vector (called the vacuum) with $D \vac = 0$ and $\cdot : V\otimes V \rightarrow V$ is a bilinear operation (called the \emph{normally ordered product}) that satisfies:
\begin{enumerate}
\item Quasi-commutativity 
\begin{equation}
a \cdot b - (-1)^{ab} b \cdot a = \int_{-\nabla}^0 \LB{a}{b} \ud\Lambda,
\label{eq:quasi-com}
\end{equation}
where the integral term is computed as follows. First compute the Lambda bracket. Second take the derivative with respect to $\chi$ to obtain an element of $V[\lambda]$. Compute the formal integral with respect to $d\lambda$. Finally evaluate the results replacing $\lambda$ by the limits: zero and $-\partial$. 
\item Quasi-associativity:
\begin{equation} \label{eq:LambdaBrackeRulesQuasiAssociativity}
                (a\cdot b)\cdot c - a\cdot (b\cdot c) = \left( \int_0^\nabla d\Lambda a \right)\cdot  [b_\Lambda c] + (-1)^{ab} \left( \int_0^\nabla d\Lambda b \right)\cdot [ a_\Lambda c]~,
                %\label{eq:2.8.2}
            \end{equation}
where the integrals are interpreted as follows: expand the Lambda bracket and put the $\Lambda$ terms inside of the integral; then take the definite integral as in the previous item. 
\item Quasi-Leibniz (non-commutative Wick formula) 
 \begin{equation}
                \LB{a}{b \cdot c }  = \LB{a}{b}\cdot  c + (-1)^{(a+1)b} b \cdot
               \LB{ a}{c} + \int_0^\Lambda \LB[\Gamma]{ \LB{a}{b} }{c} \ud \Gamma
                \label{eq:2.8.3}
            \end{equation}
\end{enumerate}
\end{defn}
From this definition we immediately see that if we rescale the Lambda bracket by $\hbar$, then the product $\cdot$ fails to be a commutative associative product satisfying Leibniz by terms of order $\hbar$. That leads naturally to the following definitions.
\begin{defn}
A SUSY Poisson vertex algebra is a tuple $(V, \vac, \cdot, \{\cdot_\Lambda \cdot\})$ where $(V, \{\cdot_\Lambda \cdot\})$ is a SUSY Lie conformal algebra, $(V,\vac, \cdot)$ is a unital associative commutative algebra, and the two operations satisfy the Leibniz rule
\begin{equation}
\{a_\Lambda b\cdot c\} = \{a_\Lambda b\} \cdot c + (-1)^{(a+1)b} a \cdot \{b_\Lambda c\}.
\label{eq:m1.7}
\end{equation}
\label{defn:poisson}
\end{defn}

\begin{defn}
A quantization of a SUSY Poisson vertex algebra $V_0$ consists of a family of SUSY vertex algebras $V_\hbar$ such that the product on $V_0$ is the limit of the product in $V_\hbar$ as $\hbar \rightarrow 0$ and the Lambda bracket of $V_0$ is given by the limit
\begin{equation}
\{a_\Lambda b\} = \lim_{\hbar \rightarrow 0} \frac{1}{\hbar} [a_\Lambda b]_\hbar.
\label{eq:m1.8}
\end{equation}
In this situation we say that $V_0$ is the \emph{quasi-classical limit} of $V_\hbar$. 
\label{defn:3}
\end{defn}

It is customary to write the Lambda bracket of a Poisson vertex algebra with braces instead of brackets mimicking the usual Poisson versus associative algebra case. 
In what follows, we will often omit the symbol $\cdot$ in products between vectors.
We will also drop the prefix SUSY.

Vertex algebras are to conformal Lie algebras what associative algebras are to Lie algebras in the following sense. For any Lie conformal algebra $\cR$ there exists a vertex algebra $U(\cR)$ with an embedding of conformal algebras $i:\cR \hookrightarrow U(\cR)$ satisfying the usual universal property: for any other vertex algebra $V$ and a map $j:\cR \rightarrow V$, there exists a morphism of vertex algebras $k:U(\cR) \rightarrow V$ such that $j = k \circ i$.  Moreover, the algebra $U(\cR)$ is constructed in very much the same way as in the Lie algebra situation, any vector of $U(\cR)$ can be obtained by products of elements of $\cR$. The notions of sub-vertex algebra and vertex algebra ideal are straightforward to define. In particular, examples \ref{ex:1}, \ref{ex:2} and \ref{ex:3} give rise to their corresponding universal enveloping vertex algebras. For any complex number $c$, we will consider their quotient by the ideal generated by $C-c$. The corresponding vertex algebras are the univeral $N=1,2$ and $4$ superconformal vertex algebras of \emph{central charge} $c$.  

This terminology justifies the notation when we say that a certain set $\mathcal{S}$ of vectors satisfying some prescribed Lambda brackets generate a vertex algebra: we first construct the corresponding Lie conformal algebra and then we consider its universal enveloping vertex algebra, in particular any vector of this algebra is a combination of products of elements of $\mathcal{S}$ and their \emph{derivatives} (\ie vectors of the $\mathcal{H}$-module generated by $\mathcal{S}$). There are other situations when the Lambda bracket of elements in $\mathcal{S}$ is not linear in the elements of $\mathcal{S}$ but can be expressed as combinations of products of elements of $\mathcal{S}$ and their derivatives. In this case we say that the vertex algebra is non-linearly generated \cite{desole}. Some of the examples  that arise as symmetries of the sigma model with target a special holonomy manifold fall into this category:
\begin{ex}[\cite{Odake:1988bh}] 
\label{ex:4}
The Odake vertex algebra is generated by two even vectors $T$ and $J$, and two odd vectors $X$ and $\bar{X}$, such that the pair $(T, J)$ generate the $N=2$ vertex algebra of central charge $c$ as in Example \ref{ex:2}, and the remaining non-trivial Lambda brackets are given by
\begin{equation}
\begin{gathered}
\begin{aligned}
{[}T_\Lambda X] &= (2\partial + \chi D + 3 \lambda) X, & [T_\Lambda \bar{X}] &= (2 \partial + \chi D + 3 \lambda) \bar{X} \\ 
[J_\Lambda X] &= -i (D + 3 \chi) X, & [J_\Lambda  \bar{X}] &= i (D + 3\chi) \bar{X}, 
\end{aligned} \\
\begin{aligned}
{[}X_\Lambda \bar{X}] &= -\frac{1}{2} \left( i TJ - (DJ) J \right) \\ & \quad  + \frac{1}{2} \chi \left( JJ - i \partial J \right) - \frac{1}{2} \lambda \left( T + i D J \right) - i \lambda \chi J - \frac{1}{2}  \lambda^2 \chi~.
\end{aligned}
\label{eq:m1.9}
\end{gathered}
\end{equation}
\end{ex}
\begin{ex}[$\spins$ algebra \cite{Shatashvili:1994zw}] This algebra is an extension of the $N=1$ algebra of central charge $12$ by an even vector $\Theta$ satisfying
\begin{equation}
	\begin{aligned}
		{[}T_\Lambda \Theta] &= \left( 2\partial + \chi D + 4 \lambda \right) \Theta +
		\frac{\chi \lambda}{2} T + \frac{2}{3} \lambda^3,\\
		[\Theta_\Lambda \Theta] &=  \frac{5}{2} \partial D \Theta + \frac{5}{4} \partial^2 T + 6 T \Theta 
		+ \\ & \quad +8 \left( \chi \partial + \lambda D + 2 \lambda \chi\right) \Theta +  \frac{15}{4} \lambda (\partial + \lambda) T
		+ \frac{8}{3} \lambda^3 \chi. 
	\end{aligned}
	\label{}
\end{equation}
\label{ex:5}
\end{ex}

\begin{ex}[$G_2$ algebra \cite{Shatashvili:1994zw}]
This vertex algebra is generated by a superconformal vector $T$ of central charge
$21/2$, an odd primary field $\Pi$ of conformal weight $3/2$, and an even field
$\Psi$ of conformal weight $2$ (but not
primary). The Lambda brackets are given by:
\begin{equation}
\bs
		{[T}_\Lambda \Psi] =& (2 \partial + \chi D + 4\lambda) \Psi -
		\frac{1}{2} \chi \lambda T - \frac{7}{12} \lambda^3, \\
\LB{\Pi}{\Pi}=& - 3 D\Psi-\frac{3}{2}\partial T - 6\chi\Psi-
 3\lambda T-\frac{7}{2}\lambda^{2}\chi ~,\\
\LB{\Pi}{\Psi}=&+3 T \Pi + \frac{5}{2}\chi\partial\Pi+ 3\lambda D\Pi+ \frac{15}{2}\chi\lambda \Pi ~,\\
\LB{\Psi}{\Psi}=& +\frac{9}{4}\partial^2T - \frac{9}{2}D\partial\Psi + 10  T 
 \Psi +  3 \Pi D \Pi
\\
& +   5 \left( \chi \partial + \lambda D + 2 \lambda \chi\right) \Psi 
\\&
+  \frac{9}{4} \lambda (\partial + \lambda) T
+\frac{35}{24}\chi\lambda^{3}  ~.\\
\end{split}
\end{equation} 
\label{ex:6}
\end{ex}
In these last three examples we see how the Lambda brackets of the generating set of vectors are expressed as quadratic expressions of the generating vectors. We remark however that the formalism of \cite{desole} is not directly applicable and in fact these algebras are not freely generated. In all these examples these algebras have been constructed by a free field realization, by exhibiting generators satisfying these OPEs. 
\def\fg{\mathfrak{g}}

Another important example we will need in this article is the $bc$--$\beta\gamma$-system:
\begin{ex}[$bc-\beta\gamma$-system] Let $V$ be a finite dimensional vector space, we will consider the algebra generated by $V\oplus V^*[1]$, that is even vectors of $V$ and odd vectors of $V^*$ with a Lambda bracket given by
\begin{equation}
[v_\Lambda w^*]_\hbar = \hbar w^*(v),
\label{eq:m1023}
\end{equation}
where $\hbar \in \mathbb{C}^\times$. These algebras are all isomorphic for any value of $\hbar$ but we include the parameter here for later convenience. 
\label{ex:6b}
\end{ex}
For completeness,  we include the example of the Boson-Fermion system:
\begin{ex}[Boson-Fermion system]
Let $W$ be a vector space with a symmetric bilinear form $(,)$. We consider the algebra generated by $W[1]$ (\ie  $W$ viewed as an odd vector space), with the Lambda bracket
\begin{equation}
[w_\Lambda w']_\hbar = \hbar \chi (w,w').
\label{eq:mboson}
\end{equation}
This is the tensor product of the Heisenberg vertex algebra with the Clifford vertex algebras associated to $W$ and $(,)$ \cite[Ex. 5.8]{Heluani:2006pk}.

In the case when $W = V \oplus V^*$ with its natural symmetric pairing, we have an embedding of the Boson-Fermion system of $W$ into the $bc$--$\beta\gamma$ system of the previous example induced by 
\begin{equation}
v \mapsto Dv, \qquad v^* \mapsto v^*.
\label{eq:membedbos}
\end{equation}
\label{ex:6c}
\end{ex}
\subsection*{State-Field correspondence} \label{sec:fields}
We include in this section an alternative definition of SUSY vertex algebras using state-superfield correspondences. This section will only be needed to explain Remark \ref{rem:superpedo} below relating the SUSY formalism with the more familiar formalism of vertex algebras in the literature. 

SUSY vertex algebras, and in particular the SUSY Lambda bracket of \cite{Heluani:2006pk} are generalizations of the more familiar concept of vertex algebras. In order to compare these we need to recall some definitions. 
   
Given a vector space $V$, an \emph{$\End(V)$-valued field} is a formal
    distribution of the form
    \begin{equation}
        A(z) = \sum_{n \in \mathbb{Z}} z^{-1-n} A_{(n)},\qquad A_{(n)} \in
        \End(V),
    \end{equation}
    such that for every $v \in V$, we have $A_{(n)}v = 0$ for large enough $n$.

\begin{defn}
    A vertex super-algebra consists of the data of a super vector space $V$,
    an even vector $\vac \in V$ (the vacuum vector),
     an even endomorphism $\partial$, and a parity preserving linear map $A \mapsto Y(A,z)$ from
     $V$ to $\End(V)$-valued fields (the state-field correspondence). This
     data should satisfy the following set of axioms:
    \begin{itemize}
    \item Vacuum axioms:
        \begin{equation}
            \begin{aligned}
            Y(\vac, z) &= \Id \\
            Y(A, z) \vac &= A + O(z) \\
            \partial \vac &= 0
            \end{aligned}
            \label{eq:1.3.1}
        \end{equation}
    \item Translation invariance:
        \begin{equation}
            \begin{aligned}
                {[}\partial, Y(A,z)] &= \partial_z Y(A,z)
        \end{aligned}
            \label{eq:1.3.2}
        \end{equation}
    \item Locality:
        \begin{equation}
            (z-w)^n [Y(A,z), Y(B,w)] = 0 \qquad n \gg 0
            \label{eq:1.3.3}
        \end{equation}
     \end{itemize}
(The notation $O(z)$ denotes a power series in $z$ without constant
term.)
    \label{defn:1.3}
\end{defn}

    Given a vertex super-algebra $V$ and a vector $A \in V$, we expand the fields
    \begin{equation}
        Y(A,z) = \sum_{{j \in \mathbb{Z}}} z^{-1-j}
        A_{(j)}
        \label{eq:1.4.1}
    \end{equation}
    and we call the endomorphisms $A_{(j)}$ the \emph{Fourier modes} of
    $Y(a,z)$. Define now the operations:
    \begin{equation}
        \begin{aligned}
            {[}A_\lambda B] &= \sum_{{j \geq 0}}
            \frac{\lambda^{j}}{j!} A_{(j)}B \\
            A B &= A_{(-1)}B
        \end{aligned}
        \label{eq:1.4.2}
    \end{equation}
    The first operation is called the $\lambda$-bracket and the second is
    called the \emph{normally ordered product}.
     The $\lambda$-bracket contains all of the information about the commutators between the Fourier coefficients of fields in $V$. 

	In order to view SUSY vertex algebras as a generalization of this state-field correspondence we have to consider \emph{super-fields} as follows. 
    We introduce formal variables $Z=(z,\theta)$ and $W =
    (w,\zeta)$, where $\theta, \zeta$ are odd
    anti-commuting variables and $z, w$ are even commuting variables.
    Given an integer $j$ and $J = 0$ or $1$ we put $Z^{j|J} = z^j \theta^J$.

    Given a super vector space $V$ and a vector $a \in V$, we will denote by
    $(-1)^a$ its parity.
    Let $U$ be a vector space, a $U$-valued formal distribution is an
    expression of the form
    \begin{equation}
        \sum_{\stackrel{j \in \mathbb{Z}}{J = 0,1}} Z^{-1-j|1-J} w_{(j|J)}
        \qquad w_{(j|J)} \in U.
        \label{eq:2.2.1}
    \end{equation}
    The space of such distributions will be denoted by $U[ [Z, Z^{-1}] ]$. If
    $U$ is a Lie algebra we will say that two such distributions $a(Z), \,
    b(W)$ are
    \emph{local} if
    \begin{equation}
        (z - w)^n [a(Z), b(W)] = 0 \qquad n \gg 0.
        \label{eq:2.2.2}
    \end{equation}
    The space of distributions such that only finitely many negative powers
    of $z$ appear (i.e. $w_{(j|J)} = 0$ for large enough $j$) will be denoted
    $U( (Z ))$. In the case when $U = \End(V)$ for another vector space $V$,
    we will say that a distribution $a(Z)$ is a \emph{field} if $a(Z)v \in V(
    (Z ))$ for all $v \in V$. The following is equivalent to Definition \ref{defn:} \cite{Heluani:2006pk}.
\begin{defn}
    An SUSY vertex algebra consists of the data of a vector space $V$,
    an even vector $\vac \in V$ (the vacuum vector), an odd endomorphism
    $D$ (whose square is an even endomorphism we denote $\partial$),
    and a parity preserving linear map $A \mapsto Y(A,Z)$ from
     $V$ to $\End(V)$-valued fields (the state-field correspondence). This
     data should satisfy the following set of axioms:
    \begin{itemize}
    \item Vacuum axioms:
        \begin{equation}
            \begin{aligned}
            Y(\vac, Z) &= \Id \\
            Y(A, Z) \vac &= A + O(Z) \\
            D \vac &= 0
            \end{aligned}
            \label{eq:2.3.1}
        \end{equation}
    \item Translation invariance:
        \begin{equation}
            \begin{aligned}
            {[} D, Y(A,Z)] &= (\partial_\theta - \theta \partial_z)
            Y(A,Z)\\
            {[}\partial, Y(A,Z)] &= \partial_z Y(A,Z)
        \end{aligned}
            \label{eq:2.3.2}
        \end{equation}
    \item Locality:
        \begin{equation}
            (z-w)^n [Y(A,Z), Y(B,W)] = 0 \qquad n \gg 0
            \label{eq:2.3.3}
        \end{equation}
     \end{itemize}
    \label{defn:2.3}
\end{defn}
\begin{rem}
    Given the vacuum axiom for a SUSY vertex algebra, we will use the state
    field correspondence to identify a vector $A \in V$ with its corresponding
    field $Y(A,Z)$.
    \label{rem:nosenosenose}
\end{rem}
    Given a $N=1$ SUSY vertex algebra $V$ and a vector $A \in V$, we expand the fields
    \begin{equation}
        Y(A,Z) = \sum_{\stackrel{j \in \mathbb{Z}}{J = 0,1}} Z^{-1-j|1-J}
        A_{(j|J)}
        \label{eq:2.4.1}
    \end{equation}
    and we call the endomorphisms $A_{(j|J)}$ the \emph{Fourier modes} of
    $Y(A,Z)$. Define now the operations:
    \begin{equation}
        \begin{aligned}
            {[}A_\Lambda B] &= \sum_{\stackrel{j \geq 0}{J = 0,1}}
            \frac{\Lambda^{j|J}}{j!} A_{(j|J)}B \\
            A B &= A_{(-1|1)}B
        \end{aligned}
        \label{eq:2.4.2}
    \end{equation}
    The first operation is called the $\Lambda$-bracket and the second is
    called the \emph{normally ordered product}.
    \label{no:2.4}

\begin{rem}
A SUSY vertex algebra $V$ with state-superfield correspondence $Y^{\mathrm{super}}$ can be viewed as a usual vertex algebra by defining the state-field correspondence $Y$ as 
\[ Y(a,z) := Y^{\mathrm{super}}(a,z,0), \qquad a \in V. \]
The operator $D$ \emph{generates supersymmetry} in the sense that we recover the superfield associated to $a$ from the usual fields associated to $a$ and $Da$:
\[ Y^{\mathrm{super}} (a,z,\theta) = Y(a,z) + \theta Y(Da,z). \]
\label{rem:super-state}
\end{rem}

\subsection*{Quasi-classical limits} \label{sec:quasi}
In the remainder of this section we describe the Poisson vertex algebras that arise as  quasi-classical limits of the vertex algebras of Examples \ref{ex:1}--\ref{ex:6}. 
The problem of quantizing a Poisson vertex algebra is a very difficult one. However, in the case of a universal enveloping vertex algebra of a Lie conformal algebra, we can construct a one parameter family just by rescaling the generators. This procedure also works when we have a central extension as in Example \ref{ex:1}.
For example, the $N=1$ algebra is generated by $T$. Denoting $\tilde{T} = \hbar T$ we have
\[ [\tilde{T}_\Lambda \tilde{T}]_\hbar = \hbar (2 \partial + \chi D + 3 \lambda) \tilde{T} + \hbar^2 \frac{\lambda^2 \chi}{3} c. \]
Rescaling this bracket by $\hbar^{-1}$, letting $\hbar \rightarrow 0$, and renaming our generator back to $T$, we obtain\begin{equation}
\{ T_{\Lambda}  T \} = (2 \partial + \chi D + 3 \lambda)  T. 
\label{eq:p1.1}
\end{equation}
We can proceed in the same way for the $N=2$ and $N=4$ vertex algebras. 

This procedure, however, will not work for the non-linear Examples \ref{ex:4}--\ref{ex:6}. We show here for completeness how all of the examples \ref{ex:1}--\ref{ex:6} can be included in a one parameter family of vertex algebras, describing also their quasi-classical limit. 
\begin{ex}[$N=1$] The family is given by
\begin{equation}
[T_\Lambda T]_\hbar = \hbar (2 \partial + \chi D + 3 \lambda) T +  \hbar^2 \frac{\lambda^2 \chi}{3} C.
\label{eq:m1.3b}
\end{equation}
Therefore its quasi-classical limit is generated by an odd vector $T$ satisfying \eqref{eq:p1.1}
\label{ex:7}
\end{ex}
\begin{ex}[$N=2$]
\label{ex:8}
The family of vertex algebras is given by \eqref{eq:m1.3b}, and the remaining brackets are:
\begin{equation}
\begin{aligned}
{[T}_\Lambda J]_\hbar &= \hbar(2\partial+ \chi D + 2\lambda) J, \\
{[J}_\Lambda J]_\hbar &= -  \hbar T - \hbar^2  \frac{C}{3} \lambda\chi~.
\end{aligned}
\label{eq:m1.4b}
\end{equation}
The corresponding Poisson vertex algebra is generated by $T$, satisfying \eqref{eq:p1.1}, and an even $J$, primary of conformal weight $1$, and with the remaining bracket
\begin{equation}
\{J_\Lambda J\} = - T.
\label{eq:p1.3}
\end{equation}
\end{ex}
\begin{ex}[$N=4$] 
The example of $N=4$ is just as in the vertex algebra situation: we have three even vectors $\{J^i\}$ and an odd vector $T$. Each pair $(T, J^i)$ gives rise to the classical limit of the $N=2$ algebra as in the previous example, and the remaining bracket is 
\[ \{J^i_\Lambda J^j\} = \varepsilon_{ijk} (D + 2 \chi) J^k. \]
\label{ex:9}
\end{ex}
\begin{ex}[The Odake algebra]
We can extend the Odake algebra to a one parameter family of vertex algebras, by considering, in addition to the family of $N=2$ algebras described in Exercise \ref{ex:7}, two vectors $X$ and $\bar{X}$ primary of conformal weight $3/2$ and the remaining brackets given by
\begin{equation}
\begin{gathered}
{[}J_\Lambda X]_\hbar = -i \hbar(D + 3 \chi) X,  \qquad  [J_\Lambda  \bar{X}] = i \hbar (D + 3\chi) \bar{X}, \\
\begin{aligned}
{[}X_\Lambda \bar{X}]_\hbar &= -\frac{1}{2} (\hbar i TJ + \hbar (DJ) J)   + \frac{1}{2} \chi \left( \hbar JJ - i \hbar^2 \partial J \right) \\ &\quad  - \frac{1}{2} \hbar^2 \lambda \left( T + i D J \right) - i \hbar^2  \lambda \chi J - \frac{1}{2}  \hbar^3 \lambda^2 \chi~.
\end{aligned}
\label{eq:m1.9b}
\end{gathered}
\end{equation}
Its quasi-classical limit is therefore generated by $T$ and $J$ as in Example \ref{ex:8}, and $X$ and $\bar{X}$ with the remaining brackets:
\begin{equation}
\begin{aligned}
\{J_\Lambda X\} &= -i (D + 3 \chi) X, &  \{J_\Lambda \bar{X} \} &= i (D + 3 \chi) \bar{X} ~,\\ 
\{X_\Lambda \bar{X}\} &= -\frac{1}{2} \left( iTJ+   (DJ)J - \chi JJ  \right)~.
\end{aligned}
\label{eq:m1.19v}
\end{equation}
\label{ex:10}
\end{ex}

\begin{ex}[$\spins$] 
\label{ex:11}
This algebra fits in the family:
\begin{equation}
	\begin{aligned}
		{[}T_\Lambda \Theta]_\hbar &= \hbar \left( 2\partial + \chi D + 4 \lambda \right) \Theta +
	\hbar^2	\frac{\chi \lambda}{2} T + \hbar^3 \frac{2}{3} \lambda^3,\\
		[\Theta_\Lambda \Theta]_\hbar &= \hbar^2 \frac{5}{2} \partial D \Theta + \hbar^3 \frac{5}{4} \partial^2 T +  6 \hbar T \Theta \\
		 & \quad +8 \hbar^2 \left( \chi \partial + \lambda D + 2 \lambda \chi\right) \Theta +  \hbar^3 \frac{15}{4} \lambda (\partial + \lambda) T
		+ \hbar^4 \frac{8}{3} \lambda^3 \chi. 
	\end{aligned}
	\label{eq:m1.19vb}
\end{equation}
Its quasi-classical limit is thus given by $T$ as in Example \ref{ex:7} and an even vector $\Theta$ satisfying:
\begin{align}
\label{eq:mnose}
\{T_\Lambda \Theta\} &= (2 \partial + \chi D + 4 \lambda) \Theta, &
\{\Theta_\Lambda \Theta\} &= 6 T \Theta.
\end{align}
\end{ex}
\begin{ex}[$G_2$]
\label{ex:12}
The family in this case is given by
\begin{equation}
\bs
\LB{\Pi}{\Pi}_\hbar=& - 3\hbar D\Psi-\hbar^2\frac{3}{2}\partial T - 6\hbar\chi\Psi-
 \hbar^{2}3\lambda T-\hbar^{3}\frac{7}{2}\lambda^{2}\chi ~,\\
\LB{\Pi}{\Psi}_\hbar=&+3\hbar T \Pi + \hbar^2\frac{5}{2}\chi\partial\Pi+\hbar^2 3\lambda D\Pi+ \hbar^{3}\frac{15}{2}\chi\lambda \Pi ~,\\
\LB{\Psi}{\Psi}_\hbar=& +\hbar^{3}\frac{9}{4}\partial^2T - \hbar^2\frac{9}{2}D\partial\Psi + 10 \hbar T 
 \Psi +  3\hbar \Pi D \Pi
\\
& + \hbar^{2}  5 \left( \chi \partial + \lambda D + 2 \lambda \chi\right) \Psi 
\\&
+  \hbar^{3}\frac{9}{4} \lambda (\partial + \lambda) T
+\hbar^{4}\frac{35}{24}\chi\lambda^{3}  ~,\\
\end{split}
\end{equation} 
from where the quasi-classical limit is given by $T$ as in Example \ref{ex:7}, two primary vectors $\Pi$ of conformal weight $3/2$, and $\Psi$ of conformal weight $2$, with the remaining brackets:
\begin{align}
\{\Pi_\Lambda \Pi\} &= -3 D \Psi, & \{\Pi_\Lambda \Psi\} &= 3 T\Pi ,&
\{\Psi_\Lambda \Psi\} &= 10 T\Psi + 3 \Pi D\Pi .
\label{eq:noasemas}
\end{align}
\end{ex}

\def\cL{\mathcal L}
\section{Classical sigma model}
\label{s:classical}
In this section, we review some basic facts about the classical sigma model. Especially, we will see the connection between covariantly constant forms on the target space and symmetries of the sigma model \cite{Howe:1991vs, Howe:1991ic}. We will also show how to write the model in the Hamiltonian framework  \cite{Zabzine:2005qf, Bredthauer:2006hf, Zabzine:2006uz}.

The position of a closed string moving on a manifold $M$ is represented by a point in the loop space $\cL M$. The corresponding phase space is given by the cotangent bundle $T^* \cL M$. This infinite dimensional manifold carries a symplectic structure and therefore we expect to have a Poisson algebra as its algebra of functions. There are many technical difficulties in identifying the right set of ``functions'' on $T^*\cL M$ to make the previous sentence rigorous. In the physics literature, typically one works with local functionals on $T^* \cL M$ effectively using algebraic variational calculus,  and the corresponding Poisson structure is in fact a Poisson vertex algebra. For an extensive review of the theory of Poisson vertex algebras and their connections with algebras of local functionals we refer the reader to \cite{barakat}. 

In the next section, we will use this connection to show how the classical symmetries of the sigma model implies the existence of certain Poisson vertex subalgebras of the quasi-classical limit of CDR. In particular,
associated to each covariantly constant form on the manifold $M$, we have corresponding local functionals on $T^* \cL M$, explicit expressions are derived in \ref{sec:currents}. In later sections we will show that they close under the Lambda bracket. 

The reader might wish to skip this section on a first-time reading since it is not needed for the main results of this article in the quantum setup of Section \ref{CDRalgExt}. 
\subsection{The sigma model in the Lagrangian formalism}
Consider the N=(1,1) supersymmetric sigma model defined on $\Sigma = S^{1}\times\mathbb{R}$. Its action is given by
\beq
S=\frac{1}{2}\int_{\Sigma}{d\sigma dt d\theta^{-}d\theta^{+} ~g_{ij}(\Phi)D_{+}\Phi^{i}D_{-}\Phi^{j}}.
\eeq{model}
We use N=(1,1) superfields $\Phi^{i}(\sigma,t,\theta^+,\theta^-)$. The circle $S^{1}$  is parametrized by $\sigma$,  and $t$, the ``time'', is the coordinate on $\mathbb{R}$. The pair $\theta^\pm$ labels the spinor coordinates, that is, we consider the supermanifold $\Sigma^{2,2}$
%:=T^*[1]\Sigma$ obtained by changing the parity of the fiber directions on $T^*\Sigma$
where $\theta^\pm$ transform as sections of the square root of the canonical bundle over $\Sigma$. 
The fields $\Phi^{i}$ are maps from $\Sigma^{2,2}$ into a target manifold $M$.
The tensor $g_{ij}$ is a given metric on $M$. The odd derivatives $D_{\pm}$ and the even derivatives $\partial_\pm$ are defined by\footnote{We here misuse spinor notation, commonly used in the physics literature. For example, the partial derivative $\partial_+$ should be understood as $\partial_{++}$ in spinor indices.}
\begin{align}
 D_\pm &= \frac{\d}{\d \theta^\pm} + \theta^\pm (\d_0 \pm \d_1)~,&
 \partial_\pm = D^2_\pm &= \d_0 \pm \d_1 ~,
\end{align}
where $\partial_0\equiv\frac{\partial}{\partial t}$ and $\partial_1\equiv\frac{\partial}{\partial \sigma}$. The equation of motion that our fields $\Phi^i$ satisfies is derived from this action as: 
\begin{equation} \label{eom}
	D_{-}D_{+}\Phi^{i}+\G^{i}_{jk}D_{-}\Phi^{j}D_{+}\Phi^{k}=0 ~,
\end{equation}
where $\Gamma$ is the Levi-Civita connection.
The model has an $N=(1,1)$ superconformal symmetry, with the corresponding current given by
\begin{equation}\label{eq:LagrTpmDef}
T_{\pm}=g_{ij}(\Phi)D_{\pm}\Phi^{i}\d_{\pm}\Phi^{j} ~.
\end{equation}
The equation of motion gives $D_{\mp} T_{\pm}=0$, that is, $T_\pm$ are conserved and also  $T_{\pm}=T_{\pm}(t \pm \sigma,\theta^\pm)$ --- we have left and right moving currents.  We can multiply $T_{\pm}$ by any function $f_{\pm}(t \pm \sigma,\theta^{\pm})$ to form $\tilde{T}_{\pm}=f_{\pm}T_{\pm}$. $\tilde{T}_{\pm}$ still satisfy  $D_{\mp}\tilde{T}_{\pm}=0$, and we therefore have infinitely many conserved currents. The components of the superfields $T_{\pm}$ are the Virasoro field and the Neveu-Schwarz supercurrent, respectively.

These are the only symmetries that we can find associated to a general Riemannian metric. However, as noticed
  in \cite{Howe:1991vs, Howe:1991ic},  if $M$ admits covariantly constant forms, the sigma model has additional symmetries. The argument goes as follows: consider a form $\omega=\omega_{i_{1}\ldots i_{n}} dx^{i_{1}}\wedge\ldots\wedge dx^{i_{n}}$ satisfying $\nabla \omega=0$, where $\nabla$ is the Levi-Civita connection. Then 
\beq
J^{(n)}_{\pm}=\omega_{i_{1}\ldots i_{n}}(\Phi)D_{\pm}\Phi^{i_1}\ldots D_{\pm}\Phi^{i_n}
\eeq{LagrSymCurrDef}
satisfies $D_{\mp}J_{\pm}=0$ on-shell, \ie with the use of $\eqref{eom}$. This implies that $J^{(n)}_{\pm}=J^{(n)}_{\pm}(t \pm \sigma,\theta^{\pm})$, and the components of $J^{(n)}_{\pm}$ will be left and right moving currents. By the same argument as above, we have infinitely many conserved currents.
The symmetries corresponding to the currents are
\beq
\delta_{\pm}\Phi^{i}=\epsilon_{\pm}g^{ii_{1}}\omega_{i_{1}\ldots i_{n}}D_{\pm}\Phi^{i_{2}}\ldots D_{\pm}\Phi^{i_{n}} ~,
\eeq{djdjk3393939}
where the parameter $\epsilon_{\pm}$ satifies $D_{\mp}\epsilon_{\pm}=0$. The action functional (\ref{model})
 is invariant under (\ref{djdjk3393939}) if $\omega$ is covariantly constant. 

\subsection{The sigma model in the Hamiltonian formalism}
 \label{classicalhamiltionian}

The sigma model $\eqref{model}$ can also be formulated in the Hamiltonian formalism \cite{Zabzine:2005qf, Bredthauer:2006hf, Zabzine:2006uz}. We integrate out one odd $\theta$, and identify the Hamiltonian and the phase space structure. In order to do so, we introduce new odd coordinates $\theta^0$ and $\theta^1$ by
 \begin{align}
  \theta^0 &= \frac{1}{\sqrt{2}} (\theta^+ + i \theta^-)~, &  \theta^1 &=\frac{1}{\sqrt{2}}(\theta^+ - i \theta^-)~,
 \end{align}
  together with odd derivatives 
 \begin{align}\label{newD}
   D_0 &= \frac{1}{\sqrt{2}} (D_+ - i D_-)~,&
   D_1 &= \frac{1}{\sqrt{2}} (D_+ + i D_-)~,
\end{align}
 which satisfy  $D^2_0 = \d_1$, $D_1^2 = \d_1$ and 
  $D_1 D_0 + D_0 D_1 = 2 \d_0$. We also  introduce  new superfields
 \begin{align} \label{newcoord}
  \phi^i &:= \Phi^i |_{\theta^0=0}~, &
  S_i &:=  g_{i j} D_0 \Phi^j |_{\theta^0=0}~,
 \end{align}
 and new derivatives
 \begin{align} 
D_1 &:= D_1|_{\theta^0=0}~, &
 \partial  &:= \partial_1~.
 \end{align}
After performing the  $\theta^0$-integration,    the action $\eqref{model}$ becomes 
 \beq 
 S= \int dt d\sigma d\theta^1 \left ( S_i \d_0 \phi^i - \frac{1}{2} H \right )~,
\eeq{actionhamiltonian}
 where
\beq
H =  \d \phi^i D_1 \phi^j g_{ij} + g^{ij} S_i D_1 S_j
 + S_k D_1 \phi^i S_l g^{jl} \Gamma^k_{~ij}~.
\eeq{}
%Since we are studying the supersymmetric case, 
The supersymmetric  sigma model phase space 
  corresponds to the cotangent bundle $T^* {\cal L}^s M$, where ${\cal L}^s M= \{S^{1|1} \rightarrow M\}$ is  the superloop space. Here $S^{1|1}$ is the supercircle $T^*[1]S^1$ with even coordinate $\sigma$ and odd coordinate $\theta^1$. This phase space is  equipped with a natural symplectic structure
\beq
  \int d\sigma\, d\theta^1 ~\delta S_i \wedge \delta \phi^i~, 
\eeq{}
where we view the fields $\phi^i$ and $S_i$ as local coordinates on $T^*\cL^s M$. 
 Thus, the space of  functionals on  $T^* {\cal L}M$ is equipped with
 a  (super) Poisson bracket $\{~,~\}$ generated 
 by the relation:  
   \beq 
   \{ \phi^i (\sigma, \theta^1), S_j (\sigma', \theta'^1) \} =
   \delta^i_j \delta (\sigma - \sigma') \delta(\theta^1 -
     \theta'^1)~.
  \eeq{poisson}
  From \eqref{actionhamiltonian},  the Hamiltonian is:  
 \beq 
 h = \frac{1}{2} \int d\sigma d\theta^1 ~H.
   \eeq{}
This Hamiltonian, together with the Poisson bracket (\ref{poisson}), generates the same dynamics as we get from the action (\ref{model}) and the variational principle. It is convenient to introduce new formal coordinates 
 on $S^{1|1}$: $\xi = e^{i\sigma}$ and $(i\xi)^{1/2} \theta = \theta^1$, which imply
 \begin{align}
 \label{newvariables123}
  (i\xi)^{1/2} D &= D_1,&
  (i\xi)^{1/2} d\theta &= d\theta^1,&
  (i\xi)^{1/2} S_{i}(\xi, \theta) &= S_i (\sigma, \theta^1).
 \end{align}
 Thus, the Poisson bracket \eqref{poisson} becomes
 \begin{equation}
 \label{newposkdkeeee}
  \{ \phi^i (\xi, \theta), S_j (\xi', \theta') \} =
   \delta^i_j~ \delta (\xi - \xi') \delta(\theta -
     \theta').
 \end{equation}
  From now on, we will use the variables $(\xi, \theta)$ on $S^{1|1}$.

\subsubsection{Poisson vertex algebras and Lambda brackets} \label{section:PVA}
The Poisson bracket between two local functionals has the following general form
\beq
\{A (\xi, \theta),  B(\xi', \theta') \}= \sum_{\stackrel{j \geq 0}{J = 0,1}}
            (-1)^J\d_{\xi'}^j D_{\xi'\theta'}^J
      \delta(\xi-\xi')\delta(\theta-\theta')  C_{(j|J)} (\xi', \theta') ~,
\eeq{eq:m123}
 where $C_{(j|J)} $ denotes the local functional multiplying the $(-1)^J\d_{\xi'}^j D_{\xi'\theta'}^J
      \delta(\xi-\xi')\delta(\theta-\theta')$-term.  Noting that the RHS of \eqref{eq:m123} looks like in \cite[Thm 4.16(4)]{Heluani:2006pk}, and in view of \cite[3.2.1.1]{Heluani:2006pk}, we denote this bracket as:
\beq
\PLB{A}{B}= \sum_{\stackrel{j \geq 0}{J = 0,1}}
            \Lambda^{j|J}C_{(j|J)} ~,
\eeq{eq:m124}
 where $\Lambda^{j|J} = \lambda^j \chi^J$ as in Section \ref{s:prelim}. The $\L$'s encode derivatives of delta functions, and the translation between the two formalisms is
\begin{equation}
    \L^{j|J} \leftrightarrow(-1)^J\d_{\xi'}^j D_{\xi'\theta'}^J
      \delta(\xi-\xi')\delta(\theta-\theta')~.
\end{equation}
For example, we write $\eqref{poisson}$ as 
\begin{equation}
\PLB{\phi^i}{S_j}=\delta^i_j ~.
\label{eq:m125}
\end{equation}
It turns out that the operation \eqref{eq:m124} satisfies the axioms of definition \ref{defn:poisson}. It is in this way we will interpret local functional computations in the classical sigma model as computations in Poisson vertex algebras.

\subsubsection{Currents in phase space} \label{sec:currents}
Next, we want to describe $T_\pm$ and $J^{(n)}_{\pm}$, given by \eqref{eq:LagrTpmDef} and \eqref{LagrSymCurrDef} respectively, in phase space coordinates. From \eqref{newD} and \eqref{newcoord} we note that
\begin{alignat}{2} 
\label{ebase}  e^{i}_{+} := D_{+}\Phi^{i}|_{\theta^{0}=0}&= \frac{\left(g^{ij}S_{j}+D\phi^{i}\right)}{\sqrt{2}}  ~,\\
\label{ebase2} i \; e^{i}_{-} := D_{-}\Phi^{i}|_{\theta^{0}=0}&=\frac{i\left(g^{ij}S_{j}-D\phi^{i}\right)}{\sqrt{2}}~.
\end{alignat}
The factor of $i$ in the definition of $e^i_-$ is introduced for later computational convenience. In terms of the set $\{ e^{i}_\pm \}$, the generators  $J^{(n)}_{\pm}$ and  $T_{\pm}$  take the form 
\beq
\bs
T_{\pm}&=\pm\left(g_{ij} D e^{i}_{\pm} e_{\pm}^{j} +g_{ij} \G^{i}_{~kl}  D\phi^{k}  e_{\pm}^{l}  e_{\pm}^{j}   \right) ~, \\
J^{(n)}_{+}&=\frac{1}{n!}\omega_{i_1\ldots i_n}e_{+}^{i_{1}}\ldots e^{i_{n}}_{+} ~,\\
J^{(n)}_{-}&=\frac{i^{n}}{n!}\omega_{i_1\ldots i_n}e_{-}^{i_{1}}\ldots e^{i_{n}}_{-}~. \\
\end{split}
\eeq{currents}
Note, that in order to find the expression for $T_\pm$ we had to use the equations of motion \eqref{eom}. 
\section{Classical algebra extensions}
\label{clAlgExt}
In the previous section we described how local functionals on the (super) loop space give rise to elements of a Poisson vertex algebra. In this section we will compute explicitly the Poisson vertex subalgebra generated by the elements \eqref{currents} associated to covariantly constant forms on $M$. All the computations in this section are straightforward applications of the Leibniz identity and the axioms of the Lambda bracket in definition \ref{defn:k.conformal.1}.

Using local coordinates $\{S_i, \phi^i\}$ (and their derivatives) for this space we have their corresponding Lambda bracket \eqref{eq:m125}. We will use the combinations $\{e^i_\pm\}$ instead and derive closed expressions for the Lambda brackets of the currents $\{J_\pm^{(n)}\}$ defined in \eqref{currents}.  

The results in this section were obtained in \cite{Howe:1991vs, Howe:1991ic}, although in a different framework. We derive them here in the language of Poisson vertex algebras.  In the next section, we interpret the formulas as a formal quantization (in the setting of vertex algebras) of the classical symmetries of the sigma model. 

Using \eqref{eq:m125} and the axioms of definition \ref{defn:poisson} we obtain the Lambda brackets between the elements \eqref{ebase}-\eqref{ebase2}:
 \begin{align}
\PLB{e^i_\pm   }{   e^{j}_\pm }&=\pm \chi g^{ij} + \frac{1}{\sqrt{2}}\left(  g^{kj}\Gamma^i_{mk}e_{\mp}^{m} -  g^{ki}\Gamma^{j}_{mk}  e_\pm^{m}   \right) ~, \\
\PLB{e^i_+    }{   e^{j}_-  }&=\frac{1}{\sqrt{2}}\left( g^{kj}\Gamma^i_{mk}e_+^{m}-g^{ki}\Gamma^{j}_{mk}e_-^{m}\right) ~,
\end{align}
and also
\begin{align}
\PLB{e^i_\pm  }{  f(\phi)  }&=\frac{1}{\sqrt{2}}g^{ij}f_{,j} ~,
\end{align}
for any smooth function $f$ on $M$,  where $f_{,j} = \partial_j f$. 

An application of the Leibniz identity to the first equation of \eqref{currents} shows that $\{T_\pm\}$ generate two commuting copies of the $N=1$ superconformal Poisson vertex algebra as in Example \ref{ex:7}, that is:
\beq
\bs
\PLB{T_{\pm}}{T_{\pm}}&=\left(2\partial+\chi D+3\lambda\right)T_{\pm} ~, \\
\PLB{T_{\mp}}{T_{\pm}}&=0 ~.
\end{split}
\eeq{eq:Wittalgebra}
Similarly another application of the Leibniz identity gives:
\beq 
\bs
\PLB{T_\pm}{J^{(n)}_\pm}&=\left(2\partial+\chi D +n\lambda\right)J^{(n)}_\pm ~,\\
\PLB{T_{\mp}}{J^{(n)}_{\pm}}&=0 ~,
\end{split}
\eeq{idontunderstandthisnumbering}
which shows that $J^{(n)}_{\pm}$ have conformal weight $\frac{n}{2}$ with respect to the left/right moving super Virasoro field. Let us define 
\beq
\bs
B^{i}_{+(n)}& = \frac{1}{n!}g^{ii_1}\omega_{i_1\ldots i_{n+1}}e^{i_{2}}_+\ldots e^{i_{n+1}}_+~,\\
B^{i}_{-(n)}& = \frac{i^{n}}{n!} g^{ii_1}\omega_{i_1\ldots i_{n+1}}e^{i_{2}}_- \ldots e^{i_{n+1}}_-~,
\end{split}
\eeq{}
by raising an index with the metric. A straightforward computation, repeatedly applying the Leibniz rule, shows:
\beq
\bs
\PLB{J^{(n)}_\pm}{J^{(m)}_\pm} &=  (-1)^{n+1}\Bigl(\chi g_{ij}B^{i}_{\pm(n-1)}B^{j}_{\pm(m-1)} \\
&\qquad +g_{ij}DB^{i}_{\pm(n-1)}B^{j}_{\pm(m-1)}+g_{ij}\Gamma^{i}_{kl}D\phi^{l}B^{k}_{\pm(n-1)}B^{j}_{\pm(m-1)}\Bigr) ~, \\
\PLB{J_{\pm}^{(n)}}{J_{\mp}^{(m)}}&=0 ~.
\end{split}
\eeq{jnbjm}
In particular, we see that the currents in the $+$ sector commute with those in the $-$ sector. 
\subsection{Currents from holonomy groups}
Our goal is to show that the currents $J_\pm^{(n)}$, together with the super Virasoro $T_\pm$, close under the Lambda bracket. To this end we need to express the RHS of \eqref{jnbjm} in terms of these elements. In order to do so we need to exploit the geometrical properties of $M$. In particular, the existence of covariantly constant forms on $M$ is reflected in the fact that the holonomy group of $M$ reduces to a subgroup of $SO(n)$. Assuming that
 $M$ is simply-connected, that the metric $g$ is irreducible (to avoid the holonomy group being a product 
      of two groups of lower dimension), and that $M$ is not locally a Riemannian symmetric space, we obtain that the holonomy group of $M$ is one of the seven possible groups in Berger's list\footnote{Berger's original list also included $\mathrm{Spin}(9)$ which was later shown to be necessarilly locally symmetric or locally flat.} (see \cite{MR1787733} for a review of the subject). 

  Below, we compute the structure of the algebra generated by $\{J^{(n)}_{\pm}\}$ and $T_\pm$ in each of these seven cases, recovering thus the examples computed in  \cite{Howe:1991vs, Howe:1991ic}. In order to compute the structure of the corresponding algebras, we will need some algebraic properties of the covariantly constant forms on special holonomy manifolds. 
   We collect the relevant formulas (most of which where derived in \cite{karigiannis-2007,karigiannis-2007-2008}) in Appendix  \ref{crosscontractions}.
\subsubsection{Orientable Riemannian manifold, \texorpdfstring{$SO(n)$}{SO(n)}}

On a general $n$-dimensional orientable Riemannian manifold the holonomy group is $SO(n)$, and we have the covariantly constant totally anti-symmetric tensor $ \epsilon_{i_1\ldots i_{n}}$. For $n > 2$ the Poisson bracket between the corresponding currents is zero. For $n=2$, because $SO(2)=U(1)$, and $\epsilon_{i_1 i_2}$ can be taken as the K\"ahler form, we get the $N=2$ supersymmetry algebra (see the next section).

\subsubsection{K\"ahler manifold, $U(n)$}

When the holonomy group is $U(n)$, $\dim M=2n$,  the manifold is K\"ahler and we have a covariantly constant $2$-form, the K\"ahler form $\omega$. Using $\omega = g I$, $I$ being the complex structure, the current is defined as 
\beq
 J_\pm^{(2)}= \pm \frac{1}{2} \omega_{ij} e^i_\pm e^j_\pm~,
\eeq{jwj39393sjjsj}
 and we find that $\eqref{jnbjm}$ reduces to
\beq
\PLB{J^{(2)}_\pm}{J^{(2)}_\pm}=-T_\pm ~.\\
\eeq{jjdjdjjdjw2220}
Therefore, when the target manifold is K\"ahler, we get two commuting copies of the N=2 superconformal Poisson vertex algebra, see Example \ref{ex:8}.

\subsubsection{Calabi-Yau, $SU(n)$}

When the holonomy group is $SU(n)$, with $\dim M=2n$, the manifold $M$ is a Calabi-Yau. We then have, in addition to the K\"ahler form,  a covariantly constant holomorphic $n$-form $\Omega$, and its complex conjugate $\bar{\Omega}$  at our disposal. Let us denote the corresponding currents $X^{(n)}_\pm$ and $\bar{X}^{(n)}_\pm$:
\begin{align}
X^{(n)}_+&=\frac{1}{n!}\Omega_{\alpha_{1}\ldots\alpha_{n}}e^{\alpha_1}_+\ldots e^{\alpha_n}_+ ~,
& X^{(n)}_- &=\frac{i^n}{n!}\Omega_{\alpha_{1}\ldots\alpha_{n}}e^{\alpha_1}_-\ldots e^{\alpha_n}_-~, \\
\bar{X}^{(n)}_+&=\frac{1}{n!}\bar{\Omega}_{\bar{\alpha}_1\ldots\bar{\alpha}_n}e^{\bar{\alpha}_1}_+\ldots e^{\bar{\alpha}_n}_+ ~,
& \bar{X}^{(n)}_-&=\frac{i^n}{n!}\bar{\Omega}_{\bar{\alpha}_1\ldots\bar{\alpha}_n}e^{\bar{\alpha}_1}_-\ldots e^{\bar{\alpha}_n}_- ~,  
\end{align}
 which are defined in addition to $J^{(2)}_\pm$ and $T_\pm$ on the Calabi-Yau manifold. 
We here introduced complex coordinates, with indices $i= ({\alpha,\bar{\alpha}})$. Choosing an Hermitian metric, we find, using the formulas in Appendix \ref{section:FormulasCalabiYau}, that $\eqref{jnbjm}$ reduces to  
\beq
\bs
\PLB{J^{(2)}_\pm}{X^{(n)}_\pm}&= - i\left(n\chi X^{(n)}_\pm+DX^{(n)}_\pm \right)~, \\
\PLB{J^{(2)}_\pm}{\bar{X}^{(n)}_\pm}&= +i \left(n\chi \bar{X}^{(n)}_\pm+D\bar{X}^{(n)}_\pm \right)~, \\
\PLB{X^{(n)}_\pm}{X^{(n)}_\pm}&=0 ~,\\
\PLB{\bar{X}^{(n)}_\pm}{\bar{X}^{(n)}_\pm}&=0 ~,\\
\PLB{X^{(n)}_\pm}{\bar{X}^{(n)}_\pm}&= \frac{i^{n^2+1}}{(n-1)!} \Biggl(\frac{i}{2} (n-1) T\left(J^{(2)}_\pm\right)^{n-2} \\
& \qquad \qquad \qquad -\frac{1}{2}D\left(J^{(2)}_\pm\right)^{n-1}-\chi \left(J^{(2)}_\pm \right)^{n-1} \Biggr)~,
\end{split}
\eeq{clasis333939s9sjj}
 where $J^{(2)}_\pm$ is defined in (\ref{jwj39393sjjsj}). Note that \eqref{idontunderstandthisnumbering} reads now
\begin{equation}
\begin{split}
\{ {T_\pm}_\Lambda X_{\pm}^{(n)} \} &= (2 \partial + \chi D + n \lambda) X_\pm^{(n)}, \\
\{ {T_\pm}_\Lambda \bar{X}_{\pm}^{(n)} \} &= (2 \partial + \chi D + n \lambda) \bar{X}_\pm^{(n)}, \\
\{ {T_\mp}_\Lambda {X}_{\pm}^{(n)} \} &= \{ {T_\mp}_\Lambda \bar{X}_{\pm}^{(n)} \} = 0.
\end{split}
\end{equation}
% These relations, together with the remaining relations for $J^{(2)}_\pm$ and $T_\pm$, give 
 % rise to two commuting copies of the classical Odake algebra, a Poisson vertex algebra that is a generalization of Example \ref{ex:10}. 
  
 In the particular case of a Calabi-Yau threefold, that is when $n=3$, this algebra reduces to the Odake algebra of Example \ref{ex:10}.
  
  Notice that the currents $(J^{(2)}_\pm, T_\pm, X^{(n)}_\pm, \bar{X}^{(n)}_\pm)$ satisfy extra constraints. For example, from $\omega \wedge \Omega= \omega \wedge \bar{\Omega}=0$ we obtain the identities
    \begin{align} \label{relations933is3}
        J_\pm^{(2)} X^{(n)}_\pm &= 0~,&
        J_\pm^{(2)} \bar{X}^{(n)}_\pm &=0~,
\end{align}
which in turn are needed to check the Jacobi identity for \eqref{clasis333939s9sjj}.

\subsubsection{Hyperk\"ahler manifold, $Sp(n)$}
When the holonomy group is $Sp(n)$,  $\dim M=4n$, the manifold $M$ is hyperk\"ahler.  We have three complex
 structures $I_A$, $A=1,2,3$, such that $I_A I_B = - \delta_{AB} + \epsilon_{ABC} I_C$. 
  The metric $g$ is Hermitian with respect to all $I_A$ and the forms $\omega_A = gI_A$ are covariantly constant.
For $\omega_A$ we denote the corresponding currents $J_{\pm A}^{(2)}$, $A=1,2,3$, where we use (\ref{jwj39393sjjsj}). 
   Equation $\eqref{jnbjm}$ reads:
\beq
\PLB{J^{(2)}_{\pm A}}{J^{(2)}_{\pm B}}=\epsilon_{ABC}\left(D+2\chi\right)J^{(2)}_{\pm C}-\delta_{AB}T_\pm~,
\eeq{hyperkahleralgebraJJ}
which, together with \eqref{eq:Wittalgebra}, show that $\{J_{\pm A}^{(2)}, T_\pm\}$ generate two commuting copies of  the  N=4 superconformal algebra of Example \ref{ex:9}.
\subsubsection{Quaternionic K\"ahler manifold, $Sp(n) \cdot Sp(1)$}
On a quaternionic K\"ahler manifold the holonomy group is $Sp(n) \cdot Sp(1)$. Locally, we have three almost 
 complex structures $J_A$ and three locally defined two-forms $\omega_A= gJ_A$. 
 Defining the covariantly constant form:
 \beq
\Sigma=\sum_{i=A}^3 \omega_A \wedge \omega_A ~,
\eeq{}
 and denoting the corresponding currents \eqref{currents} by $\Sigma_\pm$:
\beq
 \Sigma_\pm = \frac{1}{4!}~ \Sigma_{ijkl} ~e^i_\pm e^j_\pm e^k_\pm e^l_\pm,
\eeq{ssskkk}
we obtain
\beq
\PLB{\Sigma_\pm }{\Sigma_\pm}= - 4 \Sigma_\pm  T_\pm ~.
\eeq{}

\subsubsection{$G_{2}$-manifold}\label{g2sec}
$G_2$ is an example of a \emph{exceptional} holonomy group. A
$G_{2}$-manifold $M$ is seven dimensional. 
 On such a manifold there are two covariantly constant forms,  a three-form $\Pi$ and its Hodge dual $\Psi$. %I removed some statement about cross product which I think was misplaced, perhaps you wanted to say that the Hodge star is like a cross product, it'll be just a matter of moving that sentence. 
We denote the respective currents \eqref{currents} by the same letters $\Pi_\pm$ and $\Psi_\pm$: 
\beq
\Pi_+ = \frac{1}{3!} \Pi_{ijk} e^i_+ e^j_+ e^k_+~,~~~~\Pi_- = \frac{i^3}{3!} \Pi_{ijk} e^i_- e^j_- e^k_-~,~~~~\Psi_\pm= \frac{1}{4!} \Psi_{ijkl} e^i_\pm e^j_\pm e^k_\pm  e^l_\pm~.
\eeq{kskkskksk}
 Using the formulas in Appendix \ref{crosscontractions}, 
  we find that $\eqref{jnbjm}$ reduces
to
\beq
\bs  
\PLB{\Pi_\pm}{\Pi_\pm}&= -3D\Psi_\pm -6\chi\Psi_\pm ~,\\
\PLB{\Pi_\pm}{\Psi_\pm}&= 3T_\pm\Pi_\pm ~,\\
\PLB{\Psi_\pm}{\Psi_\pm}&= 10 T_\pm \Psi_\pm + 3 \Pi_\pm D \Pi_\pm ~.
\end{split}
\eeq{classicG2}
 Recall that the $4$-form 
  $\Psi$ is the Hodge dual of $\Pi$ and thus it is not independent data. At the level of currents 
   we can derive the relation
\beq
  2T_\pm \Psi_\pm + \Pi_\pm D \Pi_\pm = 0 ~.
\eeq{classicalidealG2}

\subsubsection{$\spins$-manifold}\label{spin7sec}
$\spins$ is another example of an exceptional holonomy group.
$\spins $-manifolds are eight dimensional and they admit a covariantly  constant 4-form $\Theta$ which is self-dual with respect to the Hodge involution.
  The corresponding currents $\Theta_\pm$ satisfy:
   \beq
    \Theta_\pm = \frac{1}{4!} \Theta_{ijkl} e^i_\pm e^j_\pm e^k_\pm e^l_\pm~. 
   \eeq{odo3030kddjd}
  We find that $\eqref{jnbjm}$ reduces
to
\beq
\PLB{\Theta_\pm}{\Theta_\pm}=6T_\pm\Theta_\pm ~.
\eeq{classicSpin7}

\section{The chiral de Rham complex}
\label{s:CDR}
In this section we review the construction of the chiral de Rham complex \cite{Malikov:1998dw} of a manifold $M$. Since we work explicitly in the SUSY vertex algebra formalism, we will follow \cite{heluani-2008}. The first appearance of CDR in superfield formalism goes back to \cite{benzvi-2006}.  

Let $M$ be a differentiable manifold and let us consider the bundle $E = T[1]M \oplus T^*[1]M$ on $M$ (the shifting by $1$ here means that we declare the sections of this bundle to be odd). This bundle carries a canonical symmetric pairing $E \otimes E \rightarrow C^\infty (M)$ given by
\begin{equation}
(X+\eta, Y+\zeta) =  \left( \iota_X \zeta + \iota_Y \eta \right),
\label{eq:m126}
\end{equation}
where $X$, $Y$ are vector fields and $\eta$, $\zeta$ are differential forms and $\iota_X$ is the contraction by $X$. In addition, the bundle $E$ carries a bilinear operation known as the Dorfman bracket, defined as 
\begin{equation}
[X+\eta, Y+\zeta] = [X,Y]_{\mathrm{Lie}} + \lie_X\zeta - \iota_Y d\eta,
\label{eq:dorfman}
\end{equation}
where the first term on the RHS is the Lie bracket of vector fields. This operation is not skew-symmetric, but it does satisfy the Jacobi identity.

The chiral de Rham complex of $M$ is a sheaf of SUSY vertex algebras locally generated by sections of $E$ and by smooth functions in $C^\infty(M)$. It was P. Bressler who recognized the bracket \eqref{eq:dorfman} in the construction of CDR and gave a coordinate free description of CDR using this \cite{bressler}. In the supersymmetric setting the relation between the Dorfman bracket and the Lambda bracket becomes transparent as the following theorem shows (here $\cO_M$ is the sheaf of smooth functions on $M$ and we identify the bundle $E$ with its sheaf of sections)
\begin{prop} 
There exists a sheaf $U^\mathrm{ch}(E)$ of SUSY vertex algebras on $M$ together with maps
\begin{align}
\label{eq:m128}
i : \cO_M \hookrightarrow U^\mathrm{ch}(E) , \qquad j : E \hookrightarrow U^\mathrm{ch}(E) ,
\end{align}
satisfying the following:
\begin{enumerate}
\item $i$ is a map of algebras: $i(f) i(g) = i(fg)$, 
\item $j$ imposes a relation between the Dorfman bracket on $E$ and the Lambda bracket on $U^\mathrm{ch}(E)$ : 
\[ [j(\alpha)_\Lambda j(\beta)] = j[\alpha,\beta] + 2  \chi i (\alpha, \beta) \]
for all sections $\alpha,\beta$ of $E$, 
\item $i$ and $j$ preserve the $\cO_M$-module structure of $E$, \ie $j(f\alpha) = i(f) j(\alpha)$, 
\item  the de Rham differential $d$ and  $D \in {\cal H}$ are compatible, \ie $j df = D if$,
\item  the usual commutation relation
\[ [j \alpha_\Lambda i f] =  i (\pi(\alpha) \cdot f), \] 
where $\pi :E \rightarrow TM$ is the canonical projection. 
\end{enumerate}
$U^\mathrm{ch}(E)$ is generated by $i$ and $j$ in the sense that if there exists any other triple $U',i',j'$ satisfying 1--5 above, then there exists a map of sheaves of vertex algebras $k: U^\mathrm{ch}(E) \rightarrow U'$, such that $i' = ki$ and $j' = kj$. 
\label{prop:construction}
\end{prop}

\begin{rem}
The construction of the proposition works for any \emph{Courant algebroid} $E$ (not necessarily exact) \cite{heluani-2008}. Since we will only work here with the trivial algebroid $TM\oplus T^*M$ we will call this sheaf CDR as it coincides with the sheaf defined in \cite{Malikov:1998dw}.
\label{rem:1}
\end{rem}

Although the universal construction in terms of a Courant algebroid is elegant, a more \emph{hands-on} version in terms of coordinate charts can be given as well. Below we sketch the approach of \cite{benzvi-2006}. 
On a coordinate patch of $M$ with local coordinates $\{x^i\}$ the global sections of CDR are easy to describe. For each coordinate $x^i$ we have an even section $\phi^i$ and corresponding to the vector field $\tfrac{\partial}{\partial x^i}$ we have the odd section section $S_i$. Their Lambda brackets are given by
\begin{equation}
{[\phi^i}_\Lambda S_j] = \delta^i_j, \qquad {[\phi^i}_\Lambda \phi^j] = {[S_i}_\Lambda S_j] = 0.
\label{eq:mnoase}
\end{equation}
The commutation relations \eqref{eq:mnoase} remind us of the $bc$--$\beta \gamma$-system of Example~\ref{ex:6b}, and in fact, if we were working in the algebraic category these algebras would be isomorphic. However, since we work in the smooth setting, we allow for any smooth function of the $\phi$'s while in Example \ref{ex:6b} only polynomial functions would appear. This subtlety is addressed in the original work \cite{Malikov:1998dw} (see also \cite{MR2294219} and \cite{linshaw2} for other subtleties in the smooth case with non-locally finite coverings) where CDR was constructed by first making sense of the brackets \eqref{eq:mnoase} locally and then for any change of coordinates $x^i \mapsto y^i(x^j)$ they constructed an automorphism of CDR on the intersection of the coordinate patches. This makes it possible to glue on intersections and construct a global sheaf. 

In \cite{benzvi-2006} it was observed that in the supersymmetric setting, this automorphism is easy to describe. In fact, the fields $\tilde{\phi}^i$ and $\tilde{S}_i$ associated to the coordinates $y^i$ are expressed in terms of the original fields as
\begin{equation}
\tilde{\phi}^i = y^i(\phi^j)~, \qquad \tilde{S}_i = \frac{\partial x^j}{\partial y^i}(y(\phi)) S_j~,
\label{eq:msuperch}
\end{equation}
that is, the fields $\phi^i$ transform as the coordinates do and the fields $S_i$ transform as vector fields do. 

\begin{rem}
Using the connection between SUSY vertex algebras and vertex algebras as explained in Remark \ref{rem:super-state} we see that the $2n$ generators $\{\phi^i, S_i\}$ of our SUSY vertex algebra correspond to $4n$ generators $\{\beta_i, \gamma^i, b_i, c^i\}$ when viewed as a usual vertex algebra. The corresponding fields are given by:
\begin{equation}
\begin{aligned}
Y(\gamma^i, z) &= Y^{\mathrm{super}}(\phi^i, z,0), & Y(c^i, z) &= Y^{\mathrm{super}} (D\phi^i,z,0) \\  Y(b_i, z) &= Y^{\mathrm{super}}(S_i,z,0), & Y(a_i, z) &= Y^{\mathrm{super}}(DS_i, z,0).
\end{aligned}
\label{eq:minueva}
\end{equation}
These four set of generators were introduced in the original work \cite{Malikov:1998dw}.  From \eqref{eq:msuperch} we easily see that the fields $\gamma^i$ transform as coordinates do, while the fields $b_i$ transform as vector fields do. Since $D$ is a derivation and $c^i = D\gamma^i$ it follows that $c^i$ changes as differential forms do. The fields $\beta_i$ however change in a non-tensorial manner, namely applying $D$ to both sides of the second equation in  \eqref{eq:msuperch} we obtain:
\begin{multline*} D \tilde{S}_i = D \left( \frac{\partial x^j}{ \partial y^i} (y (\phi)) \right) S_j + \frac{\partial x^j}{\partial y^i} (y(\phi)) DS_j = \\  \frac{\partial^2 x^j}{\partial y^i \partial y^k}(y(\phi)) \frac{\partial y^k}{\partial x^l} D\phi^l S_j  + \frac{\partial x^j}{\partial y^i} (y (\phi)) DS_j
\end{multline*}
which using \eqref{eq:minueva} implies that the fields $\beta_i$ transform as 
\[\tilde{\beta}_i = \frac{\partial^2 x^j}{\partial y^i \partial y^k}(y(\gamma)) \frac{\partial y^k}{\partial x^l} c^l b_j + \frac{\partial x^j}{\partial y^i}( y(\gamma)) \beta_j,  \]
showing that in fact they do not transform as vector fields due to the first term on the RHS. One of the advantages of the SUSY formalism is that the generators of CDR (the fields $\phi^i$ and $S_i$) transform as tensorial quantities as in \eqref{eq:msuperch}.
\label{rem:superpedo}
\end{rem}

Another observation is that we can obtain a family of vertex algebras by multiplying the RHS of \eqref{eq:mnoase} by $\hbar$. In this way we obtain a Poisson vertex algebra in the limit $\hbar \rightarrow 0$ as described in Section \ref{sec:quasi}. In this limit we obtain \eqref{eq:m125} and it was using this observation that CDR was proposed in \cite{Ekstrand:2009zd} as a formal canonical quantization of the classical non-linear sigma model. 

In the following sections we will construct global sections of CDR associated to differential forms of $M$.  These will be the quantum counterpart to \eqref{currents} in a very precise sense: we recover the latter from the former by taking the limit $\hbar \rightarrow 0$. After constructing these sections it would be natural to see if, in the case of covariantly constant forms, they close under the Lambda bracket, thus  obtaining a quantum counterpart to the results of the previous section. This can explicitly be carried out in the Calabi-Yau threefold case (see Theorem \ref{main:theorem}). In the general situation, however, we can only make conjectures.

\begin{rem}
As a word of caution: the multiplication in CDR is neither associative nor commutative. It is very difficult then to write down global sections. A quick look at \eqref{eq:msuperch} shows that functions and vector fields of $M$ give rise to such global sections (a fact that we already knew from Prop. \ref{prop:construction}). However, trying to construct sections of CDR from other tensors on $M$ is not trivial because of the terms on the RHS of the quasi-associativity rule \eqref{eq:LambdaBrackeRulesQuasiAssociativity} appearing under a change of coordinates. In \cite{benzvi-2006} it was noticed that one can use the Levi-Civita connection on $M$ to counteract these quasi-associativity terms in order to construct sections of CDR associated to differential two-forms. In the next section we will generalize this result to higher order forms, providing a unifying framework 
for constructing symmetry generators of the CDR. The procedure is straightforward although computationally tedious, we will write local expressions in terms of the generating fields $\phi^i, S_i$ and then check that under changes of coordinates using \eqref{eq:msuperch} these expressions are invariant. 
%to construct all the known supersymmetries of CDR. 
\label{rem:2}
\end{rem}
\section{Constructing global sections of CDR}
\label{lift}
In this section we construct the quantum analogs of \eqref{currents}. More generally, we will find an embedding of the space of differential forms $\Omega^*(M)$ into CDR. This embedding will depend on a choice of a metric and will be explicitly given in terms of the corresponding Levi-Civita connection. As mentioned in Remark \ref{rem:2}, constructing global sections of this sheaf is a subtle task due to the lack of $\cO$-module structure. The non-tensorial nature of sections of CDR manifest itself in the appearance of anomalous terms coming from quasi-associativity \eqref{eq:LambdaBrackeRulesQuasiAssociativity} under changes of coordinates. All these terms are of order of $\hbar$ and will therefore vanish at the quasi-classical limit. In particular, we will recover \eqref{currents} in this limit. 
\subsection{Constructing well-defined sections from forms} 
\label{welldefinedsections}
Let $g$ be a Riemannian metric on $M$ and let $\{x^i\}$ be a local coordinate system. We obtain a local trivialization $\{dx^i\}$ for $T^*M$ and viewing $g$ as an isomorphism $TM \simeq T^*M$ we obtain a corresponding local frame for $TM$. According to Prop \ref{prop:construction} we have the associated local sections $D \phi^i$ and $g^{ij} S_j$ of the CDR. We define the local sections $e^i_\pm$ of CDR by the same equations \eqref{ebase}--\eqref{ebase2} as in the classical case, namely: 
\begin{alignat}{2} 
\label{ebasea}  e^{i}_{+} &= \frac{\left(g^{ij}S_{j}+D\phi^{i}\right)}{\sqrt{2}}  ~,\\
\label{ebase2a} i \; e^{i}_{-} &=\frac{i\left(g^{ij}S_{j}-D\phi^{i}\right)}{\sqrt{2}}~.
\end{alignat}

Now, let $\omega \in \Omega^n(M)$ be a differential $n$-form. We work in a local coordinate system $\{x^i\}$ so that this form is locally described by
\begin{equation*}
\omega_{i_1 \dots i_n} dx^{i_1} \wedge \dots \wedge dx^{i_n}.
\end{equation*}
We define local sections of the CDR by
 the same expression as in the classical case, given by:
\beq
J^{(n)}_{+c}=\frac{1}{n!}\omega_{i_1\ldots i_n}e_{+}^{i_{1}}\ldots e^{i_{n}}_{+} ~, \qquad 
J^{(n)}_{-c}=\frac{i^{n}}{n!}\omega_{i_1\ldots i_n}e_{-}^{i_{1}}\ldots e^{i_{n}}_{-}~. 
\eeq{holamia}
We introduced the subscript ``c'', for ``classical''. Here we encounter the first subtlety. Since the multiplication in a vertex algebra is not associative, a priori we should specify an order of multiplication in expressions like \eqref{holamia}. Although, in this particular case, due to the anti-symmetry of $\omega_{i_1\ldots i_n}$, the integral terms in the RHS of \eqref{eq:LambdaBrackeRulesQuasiAssociativity} will vanish.

We will adopt the following convention: for the RHS of expressions like \eqref{holamia} we will mean
\beq
(\omega_{j_1\ldots j_n}(\phi))\left(e^{j_1} \left(e^{j_2} \left(\ldots \left(e^{j_{n-1}}e^{j_{n}}\right)\ldots \right)\right)\right),
\eeq{eq:orderes}
and we will often not write parenthesis when the order of multiplication does not alter the expression.

Let us exemplify the problem of defining global sections due to the lack of associativity of the normal ordered product. If $n=1$, that is $\omega \in \Omega^1(M)$, then the sections \eqref{holamia} correspond to well defined sections of $E = TM \oplus T^*M$ and therefore they give rise to well defined sections of CDR by Prop.\ \ref{prop:construction}. The first problem arises when $n=2$. Let $\omega=\omega_{ij}$ be a two form on $M$, and consider a change of coordinates $x^i \rightarrow y^i$. It follows easily from \eqref{eq:msuperch} that the local section $\tilde{\omega}_{ij} \tilde{e}^i_+ \tilde{e}^j_+$ (in the coordinates $y^i$) is expressed in the system of coordinates $x^i$ as % (we sum over repeated indexes)
\begin{equation}
\left(\frac{\partial{\phi}^{i'}}{\partial \tilde{\phi}^{i}} \frac{\partial {\phi}^{j'}}{\partial \tilde{\phi}^{j}} \omega_{i'j'} \right)\left( \left( \frac{\partial \tilde{\phi}^i}{\partial {\phi}^k} e^k_+ \right) \left( \frac{\partial \tilde{\phi}^j}{\partial {\phi}^l} e^l_+  \right) \right) .
\end{equation}
Here we use the shorthand  $\partial \phi^i / \partial \tilde{\phi}^j$ to mean $\tfrac{\partial x^i}{\partial y^j}( y(\phi))$ as in \eqref{eq:msuperch} to avoid cluttering. 

Using quasi-associativity and \eqref{ebasea} this reduces to 
\begin{equation}
\omega_{ij} e^i_+ e^j_+ + \hbar \frac{\partial^2 \phi^l}{\partial \tilde{\phi}^j \tilde{\phi}^m} \frac{\partial\tilde{\phi}^j}{\partial \phi^k} \frac{\partial \tilde{\phi}^m}{\partial \phi^n} g^{pk}  \omega_{lp} \partial \phi^n .
\label{minuevados}
\end{equation}
This last expression shows that the local section $J^{(2)}_{+c}$ does not define a global section of CDR . Moreover, looking at the second ``anomalous'' term, this expression hints that in order to cancel it, we may employ a connection on $TM$.  

Let us first define the local sections, where the order of the normal ordered product follows \eqref{eq:orderes}:
\beq
F_{\pm(0)} := 1, \qquad 
F_{+(k)}^{i_1\ldots i_k} :=  e_{+}^{i_1}  \dots  e_{+}^{i_k}, \qquad
F_{-(k)}^{i_1\ldots i_k} :=  i^k e_{-}^{i_1} \ldots e_{-}^{i_k}.
\eeq{eq:fdefn}
Define $G_{\pm(n,n)}=F_{\pm(n)}$ and 
for each $1\leq s \leq \lfloor \frac{n}{2} \rfloor$ we define
\begin{equation}
G^{i_{1}\ldots i_{n}}_{\pm(n,n-2 s)}=\Gamma_{k_1l_1}^{i_1}g^{i_2k_1}\partial\phi^{l_1}\ldots\Gamma_{k_{2s -1} l_{2s-1}}^{i_{2s-1}}g^{i_{2s}k_{2s-1}}\d\phi^{l_{2s-1}}F_{\pm(n-2s)}^{i_{2s+1}\ldots i_{n}},
\label{eq:gdefe}
\end{equation}
where $\Gamma^{i}_{jk}$ are the Christoffel symbols of the Levi-Civita connection associated to $g$, 
$\lfloor \cdot \rfloor$ denotes the integer part, and the second subscript denotes how many $e$'s are present in the expression.  Note that the RHS of \eqref{eq:gdefe} does not depend on the order in which the product is evaluated, this is why we do not include nested parenthesis. 

Define the numbers $T_{r,s}$ as the coefficients of the \emph{Bessel polynomials} \cite{grosswald}:
\begin{equation}
y_r  (x) = \sum_{s=0}^r T_{r,s} x^s = \sum_{s=0}^r \frac{(r + s)!}{(r-s)! s! 2^{s}} x^s,
\label{eq:bessel1}
\end{equation}
and let $T_{r,s}:=0$ when $s<0$ or $s>r$. We arrive to the main technical theorem of this article:
\begin{theorem}
Let $(M,g)$ be a Riemannian manifold. For any differential form $\omega \in \Omega^n(M)$, define 
\begin{equation}
\label{eq:edefe}
J_{\pm q} :=  \frac{1}{n!} \omega_{i_1 \dots i_{n}} E_{\pm (n)}^{i_1\dots i_n},
\end{equation} 
\begin{equation}
\label{wellDefinedSections}
E_{\pm(n)} := \sum_{s = 0}^{\lfloor \frac{n}{2} \rfloor} \hbar^s T_{n-s,s} G_{\pm(n,n-2s)}.
\end{equation}
Then $J_{\pm q}$ are well defined sections of CDR, which in the limit $\hbar \rightarrow 0$ agree with their classical counterparts \eqref{currents}.
\label{thm:3}
\end{theorem}
\begin{rem}
Theorem \ref{thm:3} produces two  embeddings 
of $\Omega^*(M)$ into global sections of CDR. 
These are clearly different from the embedding in \cite{Malikov:1998dw}. In the latter, the image of differential forms is a commutative vertex algebra (the Lambda brackets vanish). The embeddings given by Theorem \ref{thm:3} depend explicitly on the Levi-Civita connection of the metric $g$. In particular, they are only defined for a Riemannian manifold $(M, g)$. 
\label{rem:3}
\end{rem}
For simplicity we work in the plus sector and avoid the ``+'' subscripts. The minus sector is treated similarly. We first show $E_{(n)}$ satisfy a certain recursion formula and then we show the theorem by induction using this formula.
\begin{lem}
$E_{(n)}$ defined in \eqref{wellDefinedSections} satisfy the recursion formula
\begin{equation}
E_{(0)} = 1, \qquad E_{(n)}^{[i_1 \ldots i_{n}]} = e^{[i_1}  E_{(n-1)}^{i_2 \ldots i_{n}]} + \Gamma_{k l}^{[i_1}  \partial \phi^{|k} (\partial_\chi[e^{l|}_\Lambda E_{(n-1)}^{i_2 \ldots i_{n}]}])~,
\label{eq:edefe2}
\end{equation}
where $\Gamma^i_{jk}$ are the Christoffel symbols of the Levi-Civita connection and the brackets around the upper indices denotes skew-symmetrization (the vertical bars separate the indexes that are not skew-symmetrisized, so in this last expression, only the $i_j$ are skew-symmetric). 
\label{lem:1}
\end{lem}
\begin{proof}
Applying \eqref{eq:LambdaBrackeRulesQuasiAssociativity},  we note that the local sections \eqref{eq:gdefe} satisfy
\begin{gather}
\label{eq:grec}
e^{[i_1} G^{i_2 \dots i_n]}_{(n-1,n-1-2s)} = G^{[i_1 \dots i_n]}_{(n,n-2s)}, \\
\label{eq:grec2}
 \Gamma^{[i_1}_{kl} g^{i_2 | l} \partial \phi^{k|} G_{(n-2,n-2s-2)}^{i_3\dots i_n]} =  G^{[i_1 \dots i_n]}_{(n,n-2s-2)}.
\end{gather}
Indeed, let $a,b,c$ be three elements on an arbitrary SUSY vertex algebra, with $b$ even. Suppose moreover that there are no $\chi$ terms in the OPE of $a$ and $b$, that is $\partial_\chi [a_\Lambda b]=0$. Then the integral term in skew-symmetry vanishes and we have $a\cdot b = b \cdot a$. We need two applications of \eqref{eq:LambdaBrackeRulesQuasiAssociativity} in order to pass from $a(bc)$ to $b(ac)$, namely we first use quasi-associativity from $a(bc)$ to $(ab)c = (ba)c$ and then again to associate to $b(ac)$. The integral terms of \eqref{eq:LambdaBrackeRulesQuasiAssociativity} appear with different signs in these two applications, from where it follows that $a(bc) = b(ac)$.  Equation \eqref{eq:grec} is a particular case of this reasoning with $a=e^{i_1}$, $c=F_{(n-2s-1)}$ and $b$ the remaining factors in \eqref{eq:gdefe} so that $bc=G_{(n-1,n-1-2s)}$, equation \eqref{eq:grec2} is treated similarly. 

From \eqref{eq:grec} we see that the first term in \eqref{eq:edefe2} is given by
\begin{equation}
\sum_{s=0}^{\lfloor \frac{n-1}{2} \rfloor} \hbar^s T_{n-1-s,s} G_{(n,n-2s)}^{[i_1 \dots i_n]}.
\label{eq:suma1}
\end{equation}
In order to compute the second term of \eqref{eq:edefe2}, we first note from 
the definition \eqref{ebasea} 
\begin{equation}
\partial_\chi \LB{e^{i}}{e^{j}}=\hbar g^{ij}, 
	\label{eq:basechi}
\end{equation}
from where we easily obtain
\begin{equation}
\partial_\chi [e^i_\Lambda  F^{[j_1\dots j_n]}_{(n)}] = n  \hbar g^{i [j_1} F^{j_2 \dots j_n]}_{(n-1)}  
\label{eq:nala} 
\end{equation}
Recall that denoting $a=e^{i_1}$, $c=F_{(n-2s-1)}$ and $bc=G_{(n-1,n-2s-1)}$, there are no $\chi$-terms in $[a_\Lambda b]$, therefore 
the integral term in the Leibniz rule to compute $[a_\Lambda bc]$ vanish, and we obtain from \eqref{eq:nala} for $s \geq 0$:
\begin{align}
\partial_\chi\LB{e^{l}}{G^{[i_{2}\ldots i_{n}]}_{(n-1,n-2s-1)}}&= \hbar (n - 2s -1)g^{l[i_{2}}G^{i_{3}\ldots i_{n}]}_{(n-2,n-2s-2)} ~,
\end{align}
from where we can compute:
\begin{multline}
\label{almost2}
\partial_\chi\LB{e^{l}}{E_{(n-1)}^{[i_{2}\ldots i_{n}]}}= \sum_{s=0}^{\lfloor \frac{n-1}{2} \rfloor} \hbar^{s} T_{n-1-s,s} \partial_\chi {[e^l}_\Lambda G_{(n-1,n-2s-1)}^{[i_2 \dots i_n]}] = \\ \sum_{s=0}^{\lfloor \frac{n-1}{2} \rfloor} \hbar^{s+1} T_{n-1-s,s} (n-2s-1) g^{l[i_2} G^{i_3 \dots i_n]}_{(n-2,n-2s-2)},  
\end{multline}
and using \eqref{eq:grec2} the second term in \eqref{eq:edefe2} is
\begin{multline}
\label{eq:suma2}
\sum_{s=0}^{\lfloor \frac{n-1}{2} \rfloor} \hbar^{s+1} (n-2s-1)  T_{n-1-s,s} G^{[i_1\dots i_n]}_{(n,n-2s-2)} = \\ \sum_{s=1}^{\lfloor \frac{n+1}{2} \rfloor} \hbar^s (n-2s +1) T_{n-s,s-1} G^{[i_1 \dots i_n]}_{(n,n-2s)}.
\end{multline}
Recalling $T_{r,-1} = 0$ and noting that when $n$ is odd, the last term in \eqref{eq:suma2} vanishes, adding \eqref{eq:suma1} and \eqref{eq:suma2} we obtain for the RHS of \eqref{eq:edefe2}:
\begin{multline}
\sum_{s=0}^{\lfloor \frac{n}{2} \rfloor} \hbar^s \left( T_{n-1-s,s} + (n-2s + 1) T_{n-s,s-1} \right) G^{[i_1 \dots i_n]}_{(n,n-2s)} = \\ \sum_{s=0}^{\lfloor \frac{n}{2} \rfloor} \hbar^s \frac{ (n-1)!}{(n-1-2s)! s! 2^{s-1}} \left( \frac{1}{2} + \frac{s}{n-2s} \right) G_{(n,n-2s)}^{[i_1 \dots i_n]} = \\ \sum_{s=0}^{\lfloor \frac{n}{2} \rfloor} \hbar^s T_{n-s,s} G_{(n,n-2s)}^{[i_1 \dots i_n]} = E_{(n)}^{[i_1 \dots i_n]}. 
\label{eq:total}
\end{multline}
\end{proof}
\begin{proof}[Proof of Theorem \ref{thm:3}]
We need to check that under a change of coordinates the expressions for the local sections \eqref{eq:edefe} remain unchanged. 
We proceed by induction.  Let us assume that the theorem holds for $n=k-1$. 
Performing a change of coordinates using \eqref{eq:msuperch} and quasi-associativity \eqref{eq:LambdaBrackeRulesQuasiAssociativity},  we obtain:
\begin{multline}
e^{[i_1} E_{(k-1)}^{i_2 \ldots i_{k}]} = \\  \left(\frac{\partial \tilde{\phi}^{i_1}}{ \partial \phi^{a_1}} \ldots \frac{\partial \tilde{\phi}^{i_k}}{\partial \phi^{a_k}}\right) \left(e^{[a_1} E_{(k-1)}^{a_2 \ldots a_{k}]} - \frac{\partial^2 {\phi}^{[a_1}}{\partial \tilde{\phi}^{l} \partial \tilde{\phi}^m} \frac{\partial \tilde{\phi}^{|l}}{\partial {\phi}^{b}} \frac{\partial \tilde{\phi}^m}{ \partial \phi^n} \partial \phi^n \partial_\chi\left([e^{b|}_\Lambda E_{(k-1)}^{a_2 \ldots a_{k}]})] \right) \right) ~.
\label{inhomoE}
\end{multline}
This is an analog of equation \eqref{minuevados} expressing the fact that the LHS fails to transform as a differential $k$-form. Here in the LHS the fields are expressed in terms of the new coordinates $\tilde{\phi}^i = y^i (\phi)$. 
The form of the second term precisely cancels the inhomogeneous term appearing from the transformation rule of $\Gamma^{i}_{jk}$, namely:
\begin{equation}
\tilde{\Gamma}^{i}_{jk} = \frac{\partial \phi^l}{\partial \tilde{\phi}^j} \frac{\partial \phi^m}{ \partial \tilde{\phi}^k} \frac{\partial \tilde{\phi}^i}{\partial \phi^n} \Gamma^n_{lm} + \frac{\partial \tilde{\phi}^i}{\partial \phi^l} \frac{\partial^2 \phi^l}{ \partial \tilde{\phi}^j \partial \tilde{\phi}^k}.
\label{eq:tamalucocara}
\end{equation}
Using the recursion formula of Lemma \ref{lem:1}:
\begin{multline}
E_{(k)}^{[i_1 \dots i_k]} = e^{[i_1} E_{(k-1)}^{i_2 \ldots i_{k}]} +   \tilde{\Gamma}_{n l}^{[i_1}  \partial \tilde{\phi}^{|n} (\partial_\chi[e^{l|}_\Lambda E_{(k-1)}^{i_2 \ldots i_{k}]}]) = \\  \left(\frac{\partial \tilde{\phi}^{i_1}}{ \partial \phi^{a_1}} \ldots \frac{\partial \tilde{\phi}^{i_k}}{\partial \phi^{a_k}}\right) \left(e^{[a_1} E_{(k-1)}^{a_2 \ldots a_{k}]} + \Gamma_{mb}^{[a_1} \partial \phi^{|m} \partial_\chi\left([e^{b|}_\Lambda E_{(k-1)}^{a_2 \ldots a_{k}]})] \right) \right) = \\
 \left(\frac{\partial \phi^{i_1}}{ \partial \tilde{\phi}^{a_1}} \ldots \frac{\partial \phi^{i_k}}{\partial \tilde{\phi}^{a_k}}\right) E_{(k)}^{[a_1 \dots a_k]},
\end{multline}
Where in the first two terms we write the fields in terms of the new coordinates $\tilde{\phi}^i$ and in the last two in terms of the old coordinates $\phi^i$. 
Hence when multiplying both sides by $\omega_{i_i \dots i_k}$ changing coordinates as components of differential $k$-forms do, we cancel the first factor in the RHS of this last expression  
and the theorem follows. 
\end{proof}
The first few examples of \eqref{wellDefinedSections} are easy to describe. For $n=2$ we obtain
\beq
E_{(2)}^{[i j]} = F_{(2)}^{[i j]} + \hbar\Gamma_{k l}^{[i} g^{j] k} \partial \phi^l~,
\eeq{eq:E2} 
and for $n=3$ and  $n=4$ we get (in the plus sector)
\begin{equation}\label{eq:E3}
E_{(3)}^{[i_1 i_2 i_3]} =e^{[i_1} E_{(2)}^{i_2 i_3]}  + 2\hbar\Gamma_{k l}^{[i_1} g^{|k| i_2} \partial \phi^l e^{i_3]} = F_{(3)}^{[i_1 i_2 i_3]} +  3\hbar \Gamma_{k l}^{[i_1} g^{ i_2|k} \partial \phi^{l|} e^{i_3]} ~, \\
\end{equation}
and
\begin{multline}
E_{(4)}^{[i_1 i_2 i_3 i_4]} = F_{(4)}^{[i_1 i_2 i_3 i_4]} + 6 \hbar \Gamma_{k l}^{[i_1} g^{i_2|k } \partial \phi^{l|}  e^{i_3} e^{i_4]}  \\ + 3 \hbar^2 \Gamma_{k_1 l_1}^{[i_1} g^{i_2 |k_1 } \partial \phi^{l_1|}  \Gamma_{k_2 l_2}^{i_3} g^{ i_4 ]k_2} \partial \phi^{l_2}~,
\end{multline}
respectively. 
\subsection{Well-defined sections from other tensors} \label{metriclift}
In the previous subsection we constructed well defined global sections of CDR corresponding to differential forms on $M$. It is natural to ask whether the construction there can be generalized to any $(n,0)$ tensors. In particular, we would like to construct a global section corresponding to the metric tensor $g_{ij}$. From our previous analysis we see that in the limit $\hbar \rightarrow 0$, we expect to recover the classical Virasoro  \eqref{currents}. Following our prescription, we would be tempted to write locally 
\beq
g_{ij}De^{i}e^{j} + g_{ij}\G^{i}_{~kl} D\phi^k e^l e^j ~,
\eeq{metricCurrents}
with the subtlety that we must also choose an order of multiplication.

At this time we do not know whether the expression \eqref{metricCurrents} gives rise to a well defined section of CDR.  It is striking, however, that this expression, when written in complex coordinates, gives rise to a well defined section of CDR if the manifold is Calabi-Yau or hyperk\"ahler. This follows form the fact that this section can be expressed as the Lambda bracket of other  well defined sections of the form $J^{(n)}_\pm$ as above. In those cases, this section generates a copy of the $N=1$ superconformal algebra as in Example \ref{ex:1}. It would be interesting to see if this section is well defined on any Riemannian manifold.

\section{Symmetries of CDR}
\label{CDRalgExt}
In this section we study the properties of the subalgebra of CDR generated by the sections $J^{(n)}_\pm$ associated to covariantly constant forms on $M$. We show how the results of \cite{benzvi-2006,heluani-2008} in the Calabi-Yau and hyperk\"ahler case fit in the notation of this article. Moreover, we extend those results in the Calabi-Yau threefold case in our main theorem \ref{main:theorem}, and we conjecture the existence of certain symmetries of CDR in the $G_2$ case, and discuss the  $\spins$ case. In the last subsections we give some evidence for this conjecture. 
In the case when the $G_2$-manifold $\cal M$ is the product of a Calabi-Yau threefold and $S^1$, we can attach CDR to each component in the product, and construct the $G_{2}$ currents out of them from geometrical identities. This is much along the lines of \cite{FigueroaO'Farrill:1996hm}. Since we have reliably calculated the algebra on the Calabi-Yau factor, and the circle is a flat manifold, in this special case we are able to calculate the $G_2$ algebra within the CDR framework.

\subsection{N=2 algebra} 
\label{N=2Algebra} 
When $M$ is a K\"ahler manifold, we can use the K\"ahler form $\omega$ to construct two sections of CDR given by Theorem \ref{thm:3}:
\beq
J_{\pm}= \pm \frac{1}{2} \omega_{ij}e_{\pm}^{i}e_{\pm}^{j}+\frac{1}{2}\hbar ~\G^{i}_{jk}g^{jl}\omega_{il}\d\phi^{k} ~.
\eeq{84848003dmmd}
These sections coincide (modulo a sign) with the sections with the same names studied in \cite{heluani-2008}. Let us define  the local sections: 
\beq
\begin{split}
T_{+}+T_{-} &= D\phi^{i}DS_{i}+\partial\phi^{i}S_{i}-\hbar\partial D \log{\sqrt{det g_{ij}}} ~,\\
T_{+} - T_-  &=  g_{ij}D\phi^i \partial\phi^j+g^{ij}S_{i} DS_j+ \Gamma^{j}_{kl}g^{il}D\phi^{k}(S_j S_i)~,
\end{split}
\eeq{sskke030o3o3}
where we have used the K\"ahler metric $g$. It was shown in \cite{heluani-2008} that if the metric $g$ is Ricci flat, \ie $M$ is a Calabi-Yau, then the sections $J_\pm$ and $T_\pm$  satisfy the commutation relations of Example \ref{ex:8} with $c= \tfrac{3}{2} \dim M$, and $T_\pm$  define global sections. Note that the sections $J_\pm$ are well defined due to Theorem \ref{thm:3} while the sections $T_\pm$ are well defined because they arise as Lambda brackets of the well defined sections $J_\pm$ (cf. the second equation of \eqref{eq:m1.4b}).  
\subsection{The Odake algebra}
\label{OdakeAlgebra}
Let $M$ be a Calabi-Yau threefold, with the notation of the previous example, and let  $g$ be a Ricci flat metric. In addition to the K\"ahler form $\omega$ of the previous example, we have a holomorphic volume form $\Omega$ and its complex conjugate $\bar\Omega$. Let us choose holomorphic coordinates $z^\alpha$ and antiholomorphic coordinates $z^{\bar{\alpha}} = \bar{z^\alpha}$, and let $e_\pm^\alpha$ and $e_\pm^{\bar\alpha}$ be the local sections of CDR corresponding to these coordinates. Theorem \ref{thm:3} gives us in addition to the global sections $J_\pm$ and $T_\pm$ the following four sections:
\begin{align}
X_{+}&= \frac{1}{3!} \Omega_{\alpha\beta\gamma}e_{+}^{\alpha}e_{+}^{\beta}e_{+}^{\gamma} ~,
&X_{-} &= \frac{i^3}{3!} \Omega_{\alpha\beta\gamma}e_{-}^{\alpha}e_{-}^{\beta}e_{-}^{\gamma}~,\\
\bar{X}_{+}&=\frac{1}{3!} \bar{\Omega}_{\bar{\alpha}\bar{\beta}\bar{\gamma}}e_{+}^{\bar{\alpha}}e_{+}^{\bar{\beta}}e_{+}^{\bar{\gamma}} ~,
&\bar{X}_{-}&=\frac{i^3}{3!} \bar{\Omega}_{\bar{\alpha}\bar{\beta}\bar{\gamma}}e_{-}^{\bar{\alpha}}e_{-}^{\bar{\beta}}e_{-}^{\bar{\gamma}} ~.
\label{eq:seccionesextra}
\end{align}
Note that the \emph{quantum corrections} of order $\hbar$ in \eqref{eq:E3} vanish in this case since the metric is Ricci flat. 

The main theorem of this article is
\begin{theorem}
On a Calabi-Yau threefold, the two sets of sections of CDR $(T_{\pm}, J_{\pm},X_{\pm},\bar{X}_{\pm})$ given by \eqref{84848003dmmd}-\eqref{eq:seccionesextra} generate two commuting copies of the Odake algebra as in Example \ref{ex:10}. This is an extension of the two copies of the $N=2$ algebra constructed in \cite{heluani-2008}.
\label{main:theorem}
\end{theorem}
\begin{proof}
For simplicity we put $\hbar=1$. 
We may choose holomorphic coordinates $\{z^\alpha\}$ such that the holomophic volume form is constant. This implies
$\Gamma^{\alpha}_{\alpha\beta}=\Gamma^{\bar\alpha}_{\bar\alpha\bar\beta}=g^{ i j} g_{ i j ,k} =0$. 

In these coordinates, using \eqref{ebasea}-\eqref{ebase2a}, we compute:
\begin{equation} \label{kahlerbrackets}
\begin{split}
\LB{   e^{\alpha}_\pm   }{   e^\beta _\pm   }&=0 ~, \qquad {[e^{\bar\alpha}_\pm}_\Lambda e^{\bar{\beta}}_\pm] = 0\\
\LB{   e^{\alpha}_+   }{   e^\beta _-   }&=0 ~, \qquad {[e^{\bar\alpha}_+}_\Lambda e^{\bar{\beta}}_-] = 0 \\
\LB{   e^{\alpha}_\pm   }{   e^{\bar\beta}_\pm  }&=\pm \chi g^{\alpha\bar\beta} + \frac{1}{\sqrt{2}}\left(    g^{\alpha\bar\beta}_{,\bar \nu}  e_\pm^{\bar{\nu}}      -  g^{\alpha\bar\beta}_{,\nu} e_{\mp}^{\nu}         \right) ~,\\
\LB{   e^{\alpha}_+   }{   e^{\bar\beta}_-   }&=\frac{1}{\sqrt{2}}\left(   g^{\alpha\bar{\beta}}_{,\bar \nu}e_-^{\bar{\nu}}   - g^{\alpha \bar\beta}_{,\nu} e_+^{\nu}  \right)~,
\end{split}
\end{equation}
where as usual we sum over repeated indexes and $g^{\alpha \bar{\beta}}_{,\nu} := \partial_{z^{\nu}} g^{\alpha\bar{\beta}}$.

We first show that the algebra generated by $T_+$, $J_+$, $X_+$ and $\bar{X}_+$ satisfy the commutation relations of Example~\ref{ex:4}. The computation for the minus sector can be treated in a similar way. We finally prove that the plus and minus sectors commute. 
\subsubsection*{Calculation of the plus-sector}
We note from \eqref{kahlerbrackets} that we have
\begin{equation}
{[X_+}_\Lambda X_+] = {[\bar{X}_+}_\Lambda \bar{X}_+]= 0
\label{eq:zeroequal}
\end{equation}
since only $e_+$ with either holomorphic or antiholomorphic indices would occur. We now proceed to compute the commutator ${[X_+}_\Lambda \bar{X}_+]$. The computation is long, therefore we break it in different subsections.
\subsubsection{$\LB{ X }{ \Omega_{\bar\alpha\bar\beta\bar\gamma}e_+^{\bar\alpha} }$}
First, we want to calculate the bracket between $e^{\bar\alpha}_+$ and $X_+$. From \eqref{eq:quasi-com} and \eqref{kahlerbrackets} we obtain
\begin{equation}
e^{\alpha}_+ \LB{  e^{\bar\alpha}_+  }{ e^\beta _+ }  = - \LB{   e^{\bar\alpha}_+  }{  e^\beta _+} e^{\alpha}_+~.
\end{equation}
Noticing that the integral term in \eqref{eq:2.8.3} vanishes, it follows that
\begin{align}  \label{bracketepb2epep}
\LB{   e^{\bar\alpha}_+   }{    e^{[ \alpha}_+ e^{\beta ]}_+  }&= 2 \LB{ e^{\bar\alpha}_+ }{  e^{[\alpha}_+} \; e^{\beta ]}_+ ~,\\ 
\intertext{and, in general, }
\LB{   e^{\bar\alpha}_+   }{    e^{[ \alpha_1}_+ \ldots e^{\alpha_p]}_+   }&= p \LB{e^{\bar\alpha}_{+}   }{   e^{[\alpha_1}_+ } \ldots e^{\alpha_p ] }_+  ~,
\end{align}
where the brackets in the upper indices means skew-symmetrization. Another application of \eqref{eq:2.8.3} gives: 
\begin{equation}
\LB{	e^{\bar\alpha}_+   }{ X_+  } = \frac{1}{2} \chi  \Omega_{\alpha\beta\gamma}   g^{\alpha \bar\alpha} e^{\beta }_+e^{\gamma }_+   +  \frac{1}{2}  \Omega_{\alpha\beta\gamma}   \frac{  \left(     g^{\alpha \bar\alpha}_{, \nu} e_+^{\nu}       -  g^{\alpha \bar\alpha}_{,\bar\nu} e_-^{\bar\nu}        \right) }{\sqrt{2}}  e^{\beta }_+e^{\gamma }_+  ~,
\label{bracketepbX}
\end{equation}
and
\begin{equation}
\LB{\bar{ \Omega}_{\bar\alpha\bar\beta\bar\gamma} e^{\bar\alpha}_+ }{X_+} = \frac{1}{2}\Omega_{\alpha\beta\gamma}\bar{ \Omega}_{\bar\alpha\bar\beta\bar\gamma}  \left( \chi     g^{\alpha \bar\alpha}  + \frac{1}{\sqrt{2}}\left(   g^{\alpha \bar\alpha}_{, \nu} e_+^{\nu}       -  g^{\alpha \bar \alpha}_{,\bar\nu} e_-^{\bar\nu}             \right) \right)  e^{\beta }_+e^{\gamma }_+  ~.
\label{eq:c6}
\end{equation}
Using
\begin{equation}
\Omega_{\alpha\beta\gamma}g^{\alpha\bar\alpha}\bar{\Omega}_{\bar\alpha\bar\beta\bar\gamma}=g_{\beta\bar\beta}g_{\gamma\bar\gamma}-g_{\beta\bar\gamma}g_{\gamma\bar\beta}   ~,
\end{equation}
we obtain:
\begin{equation}
(\Omega_{\alpha\beta\gamma}g^{\alpha \bar\alpha}_{, \nu}   \bar{ \Omega}_{\bar\alpha\bar\beta\bar\gamma} )  e_+^{\nu}  e^{\beta }_+e^{\gamma }_+  =   \partial_\nu (g_{\beta\bar\beta}g_{\gamma\bar\gamma}-g_{\beta\bar\gamma}g_{\gamma\bar\beta} )  e_+^{\nu}  e^{\beta }_+e^{\gamma }_+  = 0.
\end{equation}
Replacing in \eqref{eq:c6}:
\begin{equation}
\LB{ \bar{\Omega}_{\bar\alpha\bar\beta\bar\gamma}e_{+}^{\bar\alpha}  }{  X_+  }= \chi g_{\beta\bar\beta}g_{\gamma\bar\gamma} \; e_+^\beta e_+^\gamma    -    \frac{1}{ \sqrt{2}} e^{\bar\nu}_- \partial_{\bar\nu} (g_{\beta\bar\beta}g_{\gamma\bar\gamma}) \; e_+^\beta e_+^\gamma  ~,
\end{equation}
and with skew-symmetry \eqref{eq:k.skew.1}:
\begin{equation} \label{eq:XOmegaepb}
\LB{ X_+ }{ \bar{\Omega}_{\bar\alpha\bar\beta\bar\gamma}e_+^{\bar\alpha} }=(\chi+D) g_{\beta\bar\beta}g_{\gamma\bar\gamma} \; e_+^\beta e_+^\gamma  +\frac{1}{\sqrt{2}} e^{\bar\nu}_- \partial_{\bar\nu}(g_{\beta\bar\beta}g_{\gamma\bar\gamma})  \; e_+^\beta e_+^\gamma  ~.
\end{equation}
\subsubsection{$\LB{   X_+  }{ \bar{ \Omega}_{\bar\alpha\bar\beta\bar\gamma}e_+^{\bar\alpha}e_+^{\bar\beta}   }$} 
From \eqref{eq:2.8.3} we obtain
\begin{equation}
\begin{split}
\LB{   X_+  }{ \bar{ \Omega}_{\bar\alpha\bar\beta\bar\gamma}e_+^{\bar\alpha}e_+^{\bar\beta}   } =& 
\LB{   X_+  }{  \bar{\Omega}_{\bar\alpha\bar\beta\bar\gamma}e_+^{\bar\alpha} }  e_+^{\bar\beta}   - e_+^{\bar\beta}  \LB{   X_+  }{  \bar{\Omega}_{\bar\alpha\bar\beta\bar\gamma}e_+^{\bar\alpha} } \\
&+ \int^\Lambda_0 \LB[\Gamma]{  \LB{   X_+  }{ \bar{ \Omega}_{\bar\alpha\bar\beta\bar\gamma}e_+^{\bar\alpha} }  }{ e_+^{\bar\beta} } \ud \Gamma    \\
=& 2 \LB{   X_+  }{ \bar{ \Omega}_{\bar\alpha\bar\beta\bar\gamma}e_+^{\bar\alpha} }  e_+^{\bar\beta} \\
&- \int^{0}_{-\Delta} \LB[\Gamma]{   e_+^{\bar\beta}  }{  \LB{   X_+  }{\bar{ \Omega}_{\bar\alpha\bar\beta\bar\gamma}e_+^{\bar\alpha} }   } \ud \Gamma \\
& + \int^\Lambda_0 \LB[\Gamma]{  \LB{   X_+  }{  \bar{\Omega}_{\bar\alpha\bar\beta\bar\gamma}e_+^{\bar\alpha} }  }{ e_+^{\bar\beta} } \ud \Gamma ~.  \\
=& 2 \LB{   X_+  }{ \bar{ \Omega}_{\bar\alpha\bar\beta\bar\gamma}e_+^{\bar\alpha} }  e_+^{\bar\beta} \\
& + \int^{\Lambda-\Delta}_0 \!\!\! \LB[\Gamma]{  \LB{   X_+  }{\bar{  \Omega}_{\bar\alpha\bar\beta\bar\gamma}e_+^{\bar\alpha} }  }{ e_+^{\bar\beta} } \ud \Gamma  ~,
\end{split}
\label{eq:c20}
\end{equation}
where in the second identity we have used \eqref{eq:quasi-com} and in the third identity we have used  $\partial_\eta \partial_\gamma \LB[\Gamma]{   e_+^{\bar\beta}  }{  \LB{   X_+  }{\bar{ \Omega}_{\bar\alpha\bar\beta\bar\gamma}e_+^{\bar\alpha} }   } = 0 $ which in turn follows easily from \eqref{eq:XOmegaepb} and the quasi-Leibniz rule \eqref{eq:2.8.3}.

Let us introduce the auxiliary combinations:
\begin{equation}
A_{\bar \beta} := g_{\alpha\bar \beta}e_+^ \alpha, \qquad 
 B_{\bar \beta} := \frac{1}{\sqrt{2}} g_{\alpha\bar\beta,\bar \sigma}e_-^{\bar \sigma}e_+^\alpha +  \uD A_{\bar \beta} ~. 
\label{eq:ayb}
\end{equation}
satisfying 
\begin{align} \label{eq:bracketsABepb}
\partial_\chi \LB{A_{\bar\gamma} }{ e_+^{\bar\beta} } &=  \delta^{\bar\beta}_{\bar\gamma}  ~, & \partial_\chi \LB{B_{\bar\gamma} }{ e_+^{\bar\beta} }  &= 0 ~.
\end{align}
We can now compute the integral term in \eqref{eq:c20} 
\begin{equation}
\chi (\lambda - \partial)(-2 A_{\bar\gamma})  +  (\lambda - \partial)(-2 B_{\bar\gamma})~,
\end{equation}
to finally obtain
\begin{equation} \label{eq:XOmegaepbepbFinal}
\begin{split}
\LB{   X_+  }{  \bar{\Omega}_{\bar\alpha\bar\beta\bar\gamma}e_+^{\bar\alpha}e_+^{\bar\beta}   } =& 2 \chi (A_{\bar\beta} A_{\bar\gamma} ) e_+^{\bar\beta}  +2  ( B_{[\bar\beta} A_{\bar\gamma]}) e_+^{\bar\beta}  \\
& +2 \partial  B_{\bar\gamma} -2  \lambda  B_{\bar\gamma}\\
& +2  \chi \partial  A_{\bar\gamma} - 2  \chi \lambda  A_{\bar\gamma} ~.
\end{split}
\end{equation}

\subsubsection{Calculation of $\LB{X_+}{\bar X_+}$}
Using the Leibniz identity \eqref{eq:2.8.3} once more, we obtain
\begin{multline} \label{eq:XXbfirst}
\LB{X_+}{\bar X_+} = \frac{1}{3!}( \LB{   X_+  }{\bar{  \Omega}_{\bar\alpha\bar\beta\bar\gamma}e_+^{\bar\alpha}e_+^{\bar\beta}   } e_+^{\bar\gamma} + (e_+^{\bar\beta}e_+^{\bar\gamma} ) \LB{   X_+  }{  \bar{\Omega}_{\bar\alpha\bar\beta\bar\gamma}e_+^{\bar\alpha}   }  \\
+ \int_0^\Gamma   \LB[\Gamma]{  \LB{   X_+  }{\bar{  \Omega}_{\bar\alpha\bar\beta\bar\gamma}e_+^{\bar\alpha}e_+^{\bar\beta}   }   }{  e_+^{\bar\gamma} } \ud \Gamma ~).
\end{multline}
The first two terms were computed above while the brackets relevant for the integral term are (we only need the first few terms because the higher order terms will not contribute to the integral)
\begin{align}
\LB{ (A_{\bar\beta} A_{\bar\gamma} ) e_+^{\bar\beta}  }{  e_+^{\bar\gamma} } &= -2 \chi A_{\bar\beta}  e_+^{\bar\beta} + 6 \chi \lambda + \ldots  ~,\\
\LB{  ( B_{[\bar\beta} A_{\bar\gamma]}) e_+^{\bar\beta}  }{ e_+^{\bar\gamma}  } &= -2 \chi B_{\bar\beta}  e_+^{\bar\beta} + \ldots~, \\
\LB{  \partial  A_{\bar\gamma} }{  e_+^{\bar\gamma}   } &= - 3 \chi \lambda  + \ldots ~,\\
\LB{   A_{\bar\gamma} }{  e_+^{\bar\gamma}   } &= + 3 \chi  + \ldots ~.
\end{align}
The integral term in \eqref{eq:XXbfirst} therefore is
\begin{multline}
  \int_0^\Gamma   \LB[\Gamma]{  \LB{   X_+  }{  \Omega_{\bar\alpha\bar\beta\bar\gamma}e_+^{\bar\alpha}e_+^{\bar\beta}   }   }{  e_+^{\bar\gamma} } \ud \Gamma = \chi ( -4 \lambda A_{\bar\beta}  e_+^{\bar\beta} + 12  \frac{\lambda^2}{2} )  
  -4 \lambda B_{\bar\beta}  e_+^{\bar\beta} \\
- 6 \chi  \frac{\lambda^2}{2}
-  6  \chi \lambda^2 
=-4 \lambda B_{\bar\beta}  e_+^{\bar\beta} -4 \chi \lambda A_{\bar\beta}  e_+^{\bar\beta}  -3 \chi \lambda^2 ~.
\end{multline}
Collecting we obtain for \eqref{eq:XXbfirst}:
\begin{equation} \label{eq:XXbarTot}
\begin{split}
\LB{X_+}{\bar X_+} &=  \frac{1}{3} \chi ((A_{\bar\beta} A_{\bar\alpha} ) e_+^{\bar\beta} )e_+^{\bar\alpha}  + \frac{1}{3} (( B_{[\bar\beta} A_{\bar\alpha]}) e_+^{\bar\beta})e_+^{\bar\alpha}   \\
& + \frac{1}{3} (\partial  B_{\bar\alpha} )e_+^{\bar\alpha} - \frac{1}{3} \lambda  B_{\bar\alpha} e_+^{\bar\alpha} \\
& + \frac{1}{3} \chi (\partial  A_{\bar\alpha})e_+^{\bar\alpha}  - \frac{1}{3}  \chi \lambda  A_{\bar\alpha} e_+^{\bar\alpha} \\
&+ \frac{1}{6}  \chi (e_+^{\bar\beta}e_+^{\bar\alpha} ) ( A_{\bar\beta} A_{\bar\alpha} )   +\frac{1}{6}  (e_+^{\bar\beta}e_+^{\bar\alpha} )( B_{[\bar\beta} A_{\bar\alpha]}  ) \\
&-\frac{2}{3} \lambda B_{\bar\beta}  e_+^{\bar\beta} -\frac{2}{3} \chi \lambda A_{\bar\beta}  e_+^{\bar\beta}  -\frac{1}{2} \chi \lambda^2 ~.
\end{split}
\end{equation}
The constant part of \eqref{eq:XXbarTot} is given by
\begin{equation}
 \frac{1}{3} (e_+^{\bar \beta}e_+^{\bar \alpha}) \left( A_{\bar\alpha} B_{\bar\beta}   \right) 
+\frac{1}{3}  \left( ( B_{\bar\alpha} A_{\bar \beta})  e_+^{\bar \alpha} - ( B_{\bar\beta} A_{\bar\alpha}) e_+^{\bar \alpha} \right) e_+^{\bar \beta}
+\frac{1}{3} \partial ( B_{\bar\beta} )  e_+^{\bar \beta} ~,
\label{XXbarConstpartEA}
\end{equation}
where we have used \eqref{eq:bracketsABepb}. Using again \eqref{eq:bracketsABepb}, the first term in \eqref{XXbarConstpartEA} can be written as
\begin{equation}
\frac{1}{3} (e_+^{\bar \beta}e_+^{\bar \alpha}) \left( A_{\bar\alpha} B_{\bar\beta}   \right) = -\frac{1}{3}  \left(  B_{\bar\beta}  A_{\bar\alpha}  \right) (e_+^{\bar \alpha}e_+^{\bar \beta}) + \frac{2}{3} \partial \left( e_+^{\bar \beta}  B_{\bar\beta}   \right) ~.
\label{XXbarConstpartEAL1}
\end{equation}
Using  quasi-associativity in the second term of \eqref{XXbarConstpartEA} we obtain:
\begin{equation}
-\frac{1}{3}  \left( (  B_{\bar\beta} A_{\bar\alpha} - B_{\bar\alpha} A_{\bar \beta})  e_+^{\bar \alpha} \right) e_+^{\bar \beta} = -\frac{2}{3}  \left(  B_{\bar\beta}  A_{\bar\alpha}  \right) (e_+^{\bar \alpha}e_+^{\bar \beta}) -\frac{2}{3} \partial (  e_+^{\bar \beta} )  B_{\bar\beta}  ~,
\label{XXbarConstpartEAL2}
\end{equation}
from where \eqref{XXbarConstpartEA} now reads
\begin{equation}
 - \left(  B_{\bar\beta}  A_{\bar\alpha}  \right) (e_+^{\bar \alpha}e_+^{\bar \beta}) 
+ \frac{2}{3} \partial \left( e_+^{\bar \beta}  B_{\bar\beta}   \right) -\frac{2}{3} \partial (  e_+^{\bar \beta} )  B_{\bar\beta} +\frac{1}{3}  e_+^{\bar \beta} \partial ( B_{\bar\beta} )  ~.
\label{XXbarConstpartEASimp}
\end{equation}
Finally, in order to recognize the generators $T_+$ and $J_+$ in this expression we apply several times quasi-associativity to
\begin{equation}
 (  B_{\bar\beta} A_{\bar\alpha}) (e_+^{\bar \alpha}e_+^{\bar \beta}) =  (B_{\bar\beta}e_+^{\bar\beta})(A_{\bar \alpha}e_+^{\bar \alpha}) +e_+^{\bar \alpha} \partial(B_{\bar\alpha})  ~,
\label{XXbarConstpartEAL3}
\end{equation}
from where \eqref{XXbarConstpartEA} is expressed as
\begin{multline}
- \left(  B_{\bar\beta} e_+^{\bar \beta} \right)   \left( A_{\bar\alpha}  e_+^{\bar \alpha}\right) 
+ \frac{2}{3} \partial \left( e_+^{\bar \beta}  B_{\bar\beta}   \right) -\frac{2}{3} \partial (  e_+^{\bar \beta} )  B_{\bar\beta} +\frac{1}{3}  e_+^{\bar \beta} \partial  B_{\bar\beta}  - e_+^{\bar \alpha} \partial B_{\bar\alpha} \\
=  -  \left(  B_{\bar\beta} e_+^{\bar \beta} \right)   \left( A_{\bar\alpha}  e_+^{\bar \alpha}\right)  
= -\frac{1}{2}  (  T_+ + i \uD J_+) i J_+  ~.
\label{XXbarConstpartEAFinal}
\end{multline}
Similarly, collecting the $\chi$-terms from \eqref{eq:XXbarTot}, 
\begin{equation} 
\frac{1}{3}  ((A_{\bar\beta} A_{\bar\alpha} ) e_+^{\bar\beta} )e_+^{\bar\alpha}   + \frac{1}{3}  (\partial  A_{\bar\alpha})e_+^{\bar\alpha} + \frac{1}{6} (e_+^{\bar\beta}e_+^{\bar\alpha} ) ( A_{\bar\beta} A_{\bar\alpha} )   ~,
\label{eq:c200}
\end{equation}
using quasi-associativity \eqref{eq:LambdaBrackeRulesQuasiAssociativity} we compute
\begin{align} 
 (e_+^{\bar\beta}e_+^{\bar\alpha} ) ( A_{\bar\beta} A_{\bar\alpha} )  &= - ( A_{\bar\beta} A_{\bar\alpha} ) (e_+^{\bar\alpha}e_+^{\bar \beta} )  + 4 \partial(  A_{\bar\alpha} e_+^{\bar\alpha}) ~, \\
 ((A_{\bar\beta} A_{\bar\alpha} ) e_+^{\bar\beta} )e_+^{\bar\alpha}  &= - ( A_{\bar\beta} A_{\bar\alpha} ) (e_+^{\bar\alpha}e_+^{\bar \beta} )  - 2 A_{\bar\alpha}  \partial e_+^{\bar\alpha} ~, \\
( A_{\bar\beta} A_{\bar\alpha} ) (e_+^{\bar\alpha}e_+^{\bar \beta} ) &=  ( A_{\bar\beta} e_+^{\bar \beta})( A_{\bar\alpha} e_+^{\bar\alpha})  + 3  \partial A_{\bar\alpha}  e_+^{\bar\alpha}   +   A_{\bar\alpha}  \partial e_+^{\bar\alpha}  ~.
\end{align}
With these we can express \eqref{eq:c200} as
\begin{equation} 
 - \frac{1}{2}  ( A_{\bar\beta} e_+^{\bar \beta})( A_{\bar\alpha} e_+^{\bar\alpha})  -\frac{1}{2}   \partial(  A_{\bar\alpha}  e_+^{\bar\alpha} )  = +\frac{1}{2}  J_+ J_+ - \frac{1}{2}  i \partial J_+  ~.
\end{equation}
The other terms in \eqref{eq:XXbarTot} are easily computed to be 
\begin{equation}
- \lambda B_{\bar \alpha} e^{\bar{\alpha}}_+ - \chi \lambda A_{\bar{\alpha}} e^{\bar\alpha}_+ - \frac{1}{2} \chi \lambda^2 = - \frac{\lambda}{2} (T_+ + i DJ_+) - \chi \lambda i J_+ - \frac{\chi \lambda^2}{2}.
\label{eq:mamamiados}
\end{equation}
Collecting together the different terms we end up with
\begin{multline}
\LB{X_+}{\bar X_+} = - \frac{1}{2}  \bigl( i T_+ J_+ - \uD J_+ J_+   - \chi J_+ J_+ +  i\chi \partial J_+  \\
 + \lambda T_+ + i \lambda \uD J_+ + 2 i \chi \lambda J_+ + \chi\lambda^{2}    \bigr ).
\end{multline}
\subsubsection{Remaining brackets}
Note that in holomorphic coordinates where the volume form is constant the second term in \eqref{84848003dmmd} vanishes. From 
\eqref{eq:2.8.3} we obtain:
\begin{equation}
\begin{split}
\LB{   e^{\gamma}_+   }{  J_+ }  &= \LB{   e^{\gamma}_+   }{   \omega_{\alpha\bar{\beta}}e_+^{\alpha}e_+^{\bar{\beta}}  }  \\
 &=\left(  \frac{    g^{\gamma\bar\gamma}\omega_{\alpha\bar{\beta},\bar\gamma}e_+^\alpha  }{\sqrt{2}}  \right)  e_+^{\bar\beta} + \omega_{\alpha\bar{\beta}} e_+^{\alpha}  \left( \chi g^{\gamma\bar{\beta}}+
 \frac{       g^{\gamma\bar\beta}_{,\bar \nu}  e_+^{\bar{\nu}}      -  g^{\gamma\bar\beta}_{,\nu} e_{-}^{\nu}           }{\sqrt{2}} \right) \\
&=i\chi e^{\gamma}_+ - i \frac{1}{\sqrt{2}}\Gamma^{\gamma}_{\alpha\beta}e^{\alpha}_+e^{\beta}_- ~, 
\end{split}
\end{equation}
where we used the fact that the volume form is constant in these coordinates to check that the corresponding integral term of \eqref{eq:2.8.3} vanishes.
Another application of the Leibniz rule gives
\begin{equation}\label{eq:LBJX}
\LB{  J_+ }{ X_+} =-i\left(3\chi+D\right)X_++\frac{3i}{\sqrt{2}}\Gamma^{\alpha}_{\mu\nu}\Omega_{\alpha\beta\gamma}e^{\mu}_+e^{\nu}_-e_+^\beta e_+^\gamma =-i\left(3\chi+D\right)X_+.
\end{equation}
Note that we have used that $\Omega$ is covariantly constant. 
The computation for ${[J_+}_\Lambda \bar{X}_+]$ is similar and we are left to check that $X_+$ and $\bar{X}_+$ are primary of conformal weight $3/2$ for $T_+$. 
This is just an application of the Jacobi identity for SUSY Lie conformal algebras \eqref{eq:k.jacobi.1} together with the second equation of \eqref{eq:m1.4} and \eqref{eq:LBJX}:
\begin{equation} 
\begin{split}
\LB{ X_+ }{T_+ }&=- \LB{X_+}{\LB[\Gamma]{J_+}{J_+}} = - \LB[\Gamma + \Lambda]{\LB{X_+}{J_+}}{J_+} -\LB[\Gamma]{J_+}{\LB{X_+}{J_+}} \\
&= - i \LB[\Gamma + \Lambda]{(3 \chi +  2 \uD)X_+}{J_+} - i \LB[\Gamma]{J_+}{(3\chi + 2\uD) X_+} \\
&= -i(-3\chi +2 (\chi +\eta) )\LB[\Gamma + \Lambda]{X_+}{J_+} -i(-3 \chi -2(\uD + \eta))\LB[\Gamma]{J_+}{X_+} \\
&= (3 \lambda + \chi \uD + 2 \partial)X_+ ~.
\end{split}
\end{equation}
The computation for $\bar{X}_+$ is similiar. We have thus proved that $T_+, J_+, X_+$ and $\bar{X}_+$ satisfy the commutation relations of the Odake algebra in Example \ref{ex:4}. 
\subsubsection{Commuting sectors}
We now turn to show that the two copies of the Odake algebra commute. We already know that the corresponding $N=2$ algebras commute \cite{heluani-2008}. It follows from the first two lines in \eqref{kahlerbrackets} that 
\begin{equation}
{[X_+}_\Lambda X_-] = {[\bar{X}_+}_\Lambda \bar{X}_-] = 0,
\label{eq:c201}
\end{equation}
since only holomorphic or antiholomorphic indices enter the equations. Finally we compute 
\begin{equation}
\LB{ \Omega_{\alpha\beta\gamma} e^\alpha_+ }{ \Omega_{\bar\alpha\bar\beta\bar\gamma}e_-^{\bar\alpha}e_-^{\bar\beta}e_-^{\bar\gamma} } = - \frac{3}{\sqrt{2}}  \Omega_{\alpha\beta\gamma} g^{\alpha \bar\alpha}_{, \delta}\Omega_{\bar\alpha\bar\beta\bar\gamma}  e^{\delta}_+e_-^{\bar\beta}e_-^{\bar\gamma}
\end{equation}
so
\begin{equation}
\begin{split}
\LB{ \bar X_- }{X_+} & \propto \Omega_{\alpha\beta\gamma} g^{\alpha \bar\alpha}_{, \delta} \Omega_{\bar\alpha\bar\beta\bar\gamma} e^{\delta}_+e_-^{\bar\beta}e_-^{\bar\gamma}e^\beta _+e^\gamma _+ \\
&= (g_{\beta \bar\beta} g_{\gamma\bar\gamma } - g_{\beta \bar\gamma } g_{\gamma \bar\beta})_{, \delta} e^{\delta}_+e^\beta _+e^\gamma _+ e_-^{\bar\beta}e_-^{\bar\gamma} \\
&=0 ~,
\end{split}
\end{equation}
and similarly for ${[X_-}_\Lambda \bar{X}_+]=0$. We compute
\begin{equation}
\LB{ J_-  }{   e^\alpha_+} \propto \Gamma^{\alpha}_{\mu\nu}e^{\mu}_+e^{\nu}_- ~,
\end{equation}
so
\begin{equation}
\LB{ J_-  }{   X_+ } \propto \Omega_{\alpha \beta \gamma}  \Gamma^{\alpha}_{\mu\nu} e^{\mu}_+e^{\nu}_- e^{\beta}_+ e^{\gamma}_+ = 0
\end{equation}
since $\Omega$ is covariantly constant. The same argument applies to $\LB{J_-  }{\bar X_+}$, $\LB{J_+  }{X_-}$, and $\LB{J_+  }{\bar X_-}$.

Finally an application of the Jacobi identity and the second equation of \eqref{eq:m1.4} we obtain $\LB{X_+}{T_-}=\LB{\bar X_+}{T_-}=\LB{X_-}{T_+}=\LB{\bar X_-}{T_+}=0$.

Thus, $(X_+,\bar{X}_+,J_+, T_+)$ commutes with  $(X_-,\bar{X}_-,J_-, T_-)$ and we proved the theorem.
\end{proof}
 \begin{rem}
As in the classical case, there are non-trivial constraints satisfied by these currents. The quantum analog of (\ref{relations933is3}) is given by
 \beq
 J_\pm X_\pm = - i \hbar \partial X_\pm~,~~~~~~~~~~~ J_\pm \bar{X}_\pm = i \hbar \partial \bar{X}_\pm~.
 \eeq{blsle00388e3}
\label{rem:2b}
\end{rem}

\subsection{$G_2$}
If $M$ is a $G_2$ manifold, we have at our disposal a three form $\Pi$ and its Hodge dual $\Psi$. Using Theorem \ref{thm:3}, we can define the sections $\Pi_\pm$ associated to the three form. We have the following conjecture:
\begin{conj}
\label{conj1}
The sections $\Pi_\pm$ generate two commuting copies of the $G_2$ algebra given in Example \ref{ex:12}. In particular, in the limit $\hbar \rightarrow 0$ we recover the classical symmetries of section \ref{g2sec}.
\end{conj}
\begin{rem} \label{remark:G2welldefqc}
From Example \ref{ex:12}, we see that the sections $\Psi_{\pm}$ and $T_{\pm}$ can be recovered from the Lambda brackets only involving $\Pi_{\pm}$. 
\end{rem}The computation in section \ref{OdakeAlgebra} was greatly simplified by the fact that one can choose local coordinates on a Calabi-Yau manifold where the holomorphic volume form is constant. This is no longer true for other special holonomy manifolds --- we cannot choose coordinates such that the components of the covariantly constant forms are constant. Moreover, in the previous example, due to the fact that the metric $g$ was Ricci flat, the $\hbar$ terms of \eqref{eq:E3} vanished. We do not know of a special coordinate system on $G_2$ manifolds that would make the quantum corrections vanish. 

There are however two examples that would give some evidence to this conjecture, that of a flat manifold, which boils down to the computation in  
  \cite{Shatashvili:1994zw,FigueroaO'Farrill:1996hm}, and that of a Calabi-Yau threefold times the circle $S^1$. Let us start with the former and choose a flat metric $g_{ij}$ and constant $\Pi_{ijk}$, with $*\Pi_{ijkl}=\Psi_{ijkl}$. We define
\beq
\bs
\Pi_+&= \frac{1}{3!} \Pi_{ijk}~e_+^ie_+^je_+^k ~,~~~~~~~~~~~~~~\Pi_-= \frac{i^3}{3!} \Pi_{ijk}~e_-^ie_-^je_-^k ~,\\
\Psi_\pm&= - \frac{1}{4!}  \Psi_{ijkl}~e_\pm^ie_\pm^j e^k_\pm e^l_\pm  \pm \hbar\frac{1}{2}g_{ij}\d e^{i}_\pm e^{j}_\pm ~.
\end{split}
\eeq{}
A computation, partly using the software presented in \cite{Ekstrand2010}, shows that these sections indeed generate two commuting copies of the algebra  of Example \ref{ex:12}.
\begin{rem}\label{remark:liftqc}
As seen above, even in the case of a flat manifold, the sections  $\Psi_{\pm}$ has quantum corrections. These quantum corrections involve the metric, and it is currently unknown how to lift them to well defined sections when $M$ is a curved manifold, c.f.\ the discussion in section \ref{metriclift}. In Conjecture \ref{conj1}, this problem is avoided due to Remark~\ref{remark:G2welldefqc}. 
\end{rem}
Using the flat realization of the tensors $\Psi$ and $\Pi$ given in \ref{localrepPiPsi}, we find the following 
 relation between currents
\beq
\hbar^{2}\frac{1}{4}\partial^2T_\pm - \hbar 2  D\partial\Psi_\pm + 2 T_\pm \Psi_\pm +   \Pi_\pm D \Pi_\pm = 0 ~,
\eeq{G_2AlgebraIdeal}
 which is the quantum version of the classical relation \eqref{classicalidealG2}.
Taking the limit $\hbar \rightarrow 0$ we see that the Howe-Papadopoulos Poisson algebras  \cite{Howe:1991vs, Howe:1991ic}   for $G_2$  are the classical limits of the  
    Shatashvili-Vafa vertex algebras  \cite{Shatashvili:1994zw}. 

We now consider a $G_2$-manifold $M = CY_{3}\times S^{1}$,  where $CY_{3}$ is a compact Calabi-Yau threefold and $S^{1}$ is a circle. This is an example of a compact $G_2$-manifold. The covariant forms are given by
\begin{equation} \label{formsG2CYS1}
\begin{split}
\Pi&=\mathop{\mathrm{Re}}(\Omega)+\omega \wedge dx^{7} ~,\\
\Psi&=\mathop{\mathrm{Im}}(\Omega)\wedge dx^{7}+\frac{1}{2}\omega\wedge \omega ~,
\end{split}
\end{equation}
where $\omega$ and $\Omega$ are the K\"ahler and holomorphic volume forms of $CY_3$, and $dx^7$ is the nonvanishing form on $S^1$. The metric $g$ on $M$ is the product of the Ricci flat metric on $CY_3$ and the flat metric on $S^1$. 

Note that we have obvious embeddings of CDR on $CY_3$ and on $S^1$ into the chiral de Rham complex of $M$. We have at our disposal the sections $T^{CY}_\pm$, $J_\pm$,$X_\pm$ and $\bar{X}_\pm$ constructed in Theorem \ref{main:theorem}. In addition, we have the sections $e^7_\pm$ corresponding to the flat coordinate $x^7$ on $S^1$. Define the following global sections of CDR on $M$: 
\begin{equation}
\begin{aligned}
T_\pm &= T^{CY}_\pm \pm e_\pm D e_\pm,\\
\Pi_\pm &=X_\pm +\bar{X}_\pm +J_+e^7_\pm ~,\\
\Psi_\pm&=\frac{J_\pm J_\pm}{2}+i (X_\pm - \bar{X}_\pm) e^7_\pm +\frac{1}{2}\hbar e^7_\pm \partial e^7_\pm ~.
\end{aligned}
\end{equation} 
We have the following proposition:
\begin{prop}
The sections $T_\pm$, $\Pi_\pm$ and $\Psi_\pm$ generate two commuting copies of the $G_2$ algebra of Example \ref{ex:12}.
\label{prop:3}
\end{prop}
\begin{proof}
We sketch here the computation in the plus sector. Denote $I_+ = -i (X_+ - \bar{X}_+)$. 
From the commutation relations of the Odake algebra we obtain:
\begin{multline}
\LB{\Pi_+}{\Pi_+}=3\hbar D\left(\frac{J_+J_+}{2}-I_+e^7_+ +\frac{1}{2}\hbar e^7_+\partial e^7_+\right) \\
+6\hbar\chi\left(\frac{J_+J_+}{2}-I_+e^7_+ +\frac{1}{2}\hbar e^7_+\partial e^7_+\right) \\
-\hbar^2\frac{3}{2}\partial\left(T_++e^7_+De^7_+\right) -\hbar^2 3\lambda\left(T_++e^7_+De^7_+\right)-\hbar^3\frac{7}{2}\chi\lambda^{2} ~,
\end{multline}
from where the first equation of \eqref{eq:noasemas} follows. 
The other equations of \eqref{eq:noasemas} follow from the Leibniz rule \eqref{eq:2.8.3}, using \eqref{blsle00388e3}, which in this context can be rewritten as:
\begin{align}
J_+I_+&=-\hbar\partial (X_+ + \bar{X}_+) ~, &
J_+ (X_+ + \bar{X}_+)&=\hbar\partial I_+ ~.
\end{align}
\end{proof}
In particular, we have proved the following proposition:
\begin{prop} There exists an embedding of the vertex algebra \cite{Shatashvili:1994zw} of Example \ref{ex:12} associated to a manifold of special holonomy $G_2$, into the tensor product of the Odake vertex algebra \cite{Odake:1988bh} of Example \ref{ex:10} associated to a Calabi-Yau threefold and 
a free Boson-Fermion system of Example \eqref{ex:6c} generated by one odd element $e$, such that $\LB{e}{e} = \chi$. 
\end{prop}

\subsection{$\spins$}
On a $\spins$ manifold we have a covariant $4$-form $\Theta$, which according to Theorem \ref{thm:3} gives rise to two global sections of CDR.
However, it is not these sections that generate the $\spins$ algebra of Example~\ref{ex:11}. 
In the case of a flat manifold, with a flat metric $g_{ij}$ and a constant $\Theta_{ijkl}$, we can define the following global section of CDR:
\beq
\Theta_\pm =  \frac{1}{4!}  \Theta_{ijkl}~e_\pm^i e_\pm^je_\pm^ke_\pm^l  \pm \hbar\frac{1}{2}g_{ij}\d e_\pm^{i}e_\pm^{j} ~.
\eeq{thetaspin7}
Using the identities of Appendix \ref{crosscontractions} it is possible to check that  $\Theta_\pm$ fulfills the commutation relations of Example~\ref{ex:11}. We here relied on \cite{Ekstrand2010}. 
We see that $\Theta_\pm$ has quantum corrections even in the case of a flat manifold, and analogously to Remark~\ref{remark:liftqc}, it is not known how to define these quantum corrections for a curved manifold.

In contrast to the $G_2$ case, we can not recover the sections  $\Theta_\pm$ from Lambda brackets of known sections, c.f.\ Remark~\ref{remark:G2welldefqc}.

%
%%, denoted by $\Theta_\pm$. It is natural to conjecture: 
%\begin{conj}
%The sections $\Theta_\pm$ generate two commuting copies of the $\spins$ algebra of Example~\ref{ex:11}.
%\label{conj2}
%\end{conj}
%\begin{rem}
%Note that in all the ``non-linear'' Examples \ref{ex:10}--\ref{ex:12} the element $T$ can be recovered from the Lambda brackets of the other generating vectors.
%\label{rem:34}
%\end{rem}

\section{Discussion}
\label{s:summary}

In this article, we studied extensions of the super-Virasoro algebra  within the framework of the chiral de Rham complex. The main result of
 this work is the construction of two commuting copies of the Odake algebra associated to any Calabi-Yau threefold. Another result of this article is the systematic construction of global sections of CDR from antisymmetric tensors on $M$. Moreover, 
  we presented the full classical and partial quantum results for general extensions of the super-Virasoro
   algebra.  The central idea behind our consideration is the interpretation of CDR as a formal canonical 
    quantization of the non-linear sigma model.
    
    The main unresolved questions are the full calculations of the algebras for $G_2$ and $\spins$ manifolds. For $G_2$, we have a well defined generator in the curved case and in principle we can generate the other relevant generators through the brackets. By construction, these will be well-defined.
     Unfortunately, at the present moment, these calculations appear to be too complicated to carry out. 

%      Moreover, in this case we do not understand how to make the stress tensor to be a well-defined section. 

One motivation for studying manifolds with special holonomies comes from string theory. After compactification, let $M$ be the internal manifold. In order to have space-time supersymmetry,  $M$ must admit a covariantly constant spinor. For different dimensions of $M$, the constraint of  space-time supersymmetry leads to different choices of holonomy groups. For dimensions 6,7 and 8 the holonomy groups are $SU(3)$, $G_{2}$ and $\spins$. In the quantum setup, the extensions of N=(2,2) symmetry algebra for $d=6$ were studied for the first time in \cite{Odake:1988bh}, whereas the cases $d=7$ and $d=8$ were first investigated in \cite{Shatashvili:1994zw}. A common feature for the calculations of these algebra extensions performed in the mentioned papers, is that they are performed in the large volume limit, which means that the metric is treated like a flat metric. With the construction of the CDR, and its interpretation as the canonical quantization of the sigma model, we can begin to compute the symmetry algebras in a reliable way without taking the large volume limit.

Finally, we point out that the main technical tool in this article is an embedding of the space of differential forms $\Omega^*(M)$ of a Riemannian manifold $M$ into global secitons of CDR (Theorem \ref{thm:3}). This embedding is motivated by sigma model
  considerations, and it would be very interesting to further study the properties of this map, in particular, we do not currently know the reason for the appearance of Bessel polynomials in \eqref{wellDefinedSections}.

%\bigskip\bigskip
%\noindent{\bf\Large Acknowledgement}:
%\bigskip\bigskip

%\noindent
\section*{Acknowledgement}
The research of R.H. was supported by NSF grant DMS-0635607002.  
The research of M.Z. was supported by VR-grant  621-2008-4273.
We wish to thank the anonymous referees for numerous comments and suggestions. 

\appendix
%\appendixpage

\section{Special holonomy manifolds}
\label{crosscontractions}

In this appendix, we collect the relevant relations of the invariant tensors on special holonomy manifolds. 

\subsection{K\"ahler manifolds}

On a K\"ahler manifold we have the K\"ahler form $\omega = g I$, where $g$ is a metric and $I$ is a complex structure. 
 We then have $\omega g^{-1} \omega = - g$. In components
\begin{equation}
\omega_{ij}g^{jk}\omega_{kl}=-g_{il} ~.
\end{equation}

\subsection{Calabi-Yau manifolds} \label{section:FormulasCalabiYau}

On a Calabi-Yau $n$-fold, we define  the holomorphic volume form $\Omega$ and its complex 
 conjugate $\bar{\Omega}$.  The following relation holds
\begin{equation} \label{holoformula}
\Omega_{\alpha_1\alpha_2\ldots\alpha_n}g^{\alpha_1\bar{\alpha}_1}\Omega_{\bar{\alpha}_1\bar{\alpha}_2\ldots \bar{\alpha}_n}= g_{\alpha_2\bar{\alpha}_2}\ldots g_{\alpha_n\bar{\alpha}_n}  + ... ~,
\end{equation}
 where dots stand for the terms required by antisymmetrization in $\alpha_1  .... \alpha_n$ and $\bar{\alpha}_1 ... \bar{\alpha}_n$.
   In particular, for a Calabi-Yau threefold we have 
 \beq
  \Omega_{\alpha_1\alpha_2\alpha_3} g^{\alpha_1\bar{\alpha}_1} \Omega_{\bar{\alpha}_1\bar{\alpha}_2\bar{\alpha}_3}=
   g_{\alpha_2\bar{\alpha}_2} g_{\alpha_3\bar{\alpha}_3} -  g_{\alpha_2\bar{\alpha}_3} g_{\alpha_3\bar{\alpha}_2}~. 
 \eeq{ksks003383}
We also have the K\"ahler form $\omega$, and the following contraction between $\Omega$ and $\omega$:
\begin{equation}
\omega_{\alpha\bar{\alpha}}g^{\bar{\alpha}\beta_1}\Omega_{\beta_1 \ldots \beta_n}=i\Omega_{\alpha\beta_2 \ldots \beta_n} ~.
\end{equation}

\subsection{$G_2$-manifolds}

On a $G_2$ manifold we have couple of forms: a three-form $\Pi$ and its dual $*\Pi\equiv \Psi$. The following formulas are collected from \cite{karigiannis-2007}.
\begin{align} 
\Pi_{ijk} \Pi_{abc} g^{kc}  &=  g_{ia} g_{jb} - g_{ib} g_{ja} - \Psi_{ijab} \label{phiphi} \\
 \Pi_{ijk} \Psi_{abcd} g^{kd} & =  g_{ia} \Pi_{jbc} + g_{ib} \Pi_{ajc} +
g_{ic} \Pi_{abj} \\ 
& \quad - g_{aj} \Pi_{ibc} - g_{bj} \Pi_{aic} - g_{cj} \Pi_{abi} \nonumber \\
\Psi_{ijkl} \Psi_{abcd} g^{ld} &=  -\Pi_{ajk} \Pi_{ibc} - \Pi_{iak} \Pi_{jbc} - \Pi_{ija} \Pi_{kbc}  + g_{ia} g_{jb} g_{kc} \\
& \quad + g_{ib} g_{jc} g_{ka} + g_{ic} g_{ja} g_{kb}   - g_{ia} g_{jc} g_{kb} - g_{ib} g_{ja} g_{kc} - g_{ic} g_{jb} g_{ka} \nonumber \\  
& \quad  -g_{ia} \Psi_{jkbc} - g_{ja} \Psi_{kibc} - g_{ka} \Psi_{ijbc}  + g_{ab} \Psi_{ijkc} - g_{ac} \Psi_{ijkb} \nonumber
\end{align}
%\subsubsection*{Contracting $\Pi$ with itself}
%\begin{align} 
%\Pi_{ijk} \Pi_{abc} g^{ia} g^{jb} g^{kc}  &=  42  \\ 
%\Pi_{ijk} \Pi_{abc} g^{jb} g^{kc} & = 6 g_{ia} \\
%\Pi_{ijk} \Pi_{abc} g^{kc}  &=  g_{ia} g_{jb} - g_{ib} g_{ja} - \Psi_{ijab} \label{phiphi}
%\end{align}
%\subsubsection*{Contracting $\Pi$ and $\Psi$}
%\begin{align}
%\Pi_{ijk} \Psi_{abcd} g^{ib} g^{jc} g^{kd} &= 0 \\
%\Pi_{ijk} \Psi_{abcd} g^{jc} g^{kd}  &=  - 4 \Pi_{iab} \\
% \Pi_{ijk} \Psi_{abcd} g^{kd} & =  g_{ia} \Pi_{jbc} + g_{ib} \Pi_{ajc} +
%g_{ic} \Pi_{abj} \\ 
%& \quad - g_{aj} \Pi_{ibc} - g_{bj} \Pi_{aic} - g_{cj} \Pi_{abi} \nonumber 
%\label{phipsi}
%\end{align}
%\subsubsection*{Contracting $\Psi$ with itself}
%\begin{align}
%\Psi_{ijkl} \Psi_{abcd} g^{ia} g^{jb} g^{kc} g^{ld}  &=  168 \\
%\Psi_{ijkl} \Psi_{abcd} g^{jb} g^{kc} g^{ld} &=  24 g_{ia} \\
%\Psi_{ijkl} \Psi_{abcd} g^{kc} g^{ld} &=  4 g_{ia} g_{jb} - 4
%g_{ib} g_{ja} - 2 \Psi_{ijab} \\
%\label{PsiPsigPipi}\Psi_{ijkl} \Psi_{abcd} g^{ld} &=  -\Pi_{ajk} \Pi_{ibc} - \Pi_{iak} \Pi_{jbc} - \Pi_{ija} \Pi_{kbc} \\
%&  {} + g_{ia} g_{jb} g_{kc} + g_{ib} g_{jc} g_{ka} + g_{ic} g_{ja} g_{kb}  \nonumber \\
%& {} - g_{ia} g_{jc} g_{kb} - g_{ib} g_{ja} g_{kc} - g_{ic} g_{jb} g_{ka} \nonumber \\  
%& {} -g_{ia} \Psi_{jkbc} - g_{ja} \Psi_{kibc} - g_{ka} \Psi_{ijbc} \nonumber\\
%& {} + g_{ab} \Psi_{ijkc} - g_{ac} \Psi_{ijkb} \nonumber
% \label{psipsi}
%\end{align}

\subsubsection*{Local representations of $\Pi$ and $\Psi$ } \label{localrepPiPsi}

In a local orthonormal and flat basis, where the metric is $\sum dx^i \otimes dx^i$, the forms  $\Pi$ and $\Psi$ can be written \cite{Shatashvili:1994zw} as
\begin{align}
\Pi =& dx^1 dx^2 dx^5 +dx^1 dx^3 dx^6 +dx^1 dx^4 dx^7 -dx^2 dx^3 dx^7+ \\
&dx^2 dx^4 dx^6 -dx^3 dx^4 dx^5 +dx^5 dx^6 dx^7 ~, \nonumber \\
\Psi =&  dx^1 dx^2 dx^3 dx^4 -  dx^1 dx^2 dx^6 dx^7  +  dx^1 dx^3 dx^5 dx^7  - dx^1 dx^4 dx^5 dx^6 \\
&  +   dx^2 dx^3 dx^5 dx^6
  + dx^2 dx^4 dx^5 dx^7 +  dx^3 dx^4 dx^6 dx^7  ~, \nonumber
\end{align}
where the product is understood as the wedge product. 

\subsection{$\spins$-manifolds}

On a $\spins$-manifold we have a self-dual $4$-form $\Theta$. The following formula is from \cite{karigiannis-2007-2008}:
%\begin{eqnarray}
%\Theta_{ijkl} \Theta_{abcd} g^{ia} g^{jb} g^{kc} g^{ld} & = & 336 \\
%\Theta_{ijkl} \Theta_{abcd} g^{jb} g^{kc} g^{ld} & = & 42 g_{ia} \\
%\Theta_{ijkl} \Theta_{abcd} g^{kc} g^{ld} & = & 6 g_{ia} g_{jb} - 6
%g_{ib} g_{ja} - 4 \Theta_{ijab} \\ 
\begin{multline}
\Theta_{ijkl} \Theta_{abcd} g^{ld}  = g_{ia} g_{jb} g_{kc} + g_{ib} g_{jc} g_{ka} + g_{ic}g_{ja} g_{kb} - g_{ia} g_{jc} g_{kb} - g_{ib} g_{ja} g_{kc} \\
- g_{ic} g_{jb} g_{ka}  -g_{ia} \Theta_{jkbc} - g_{ja} \Theta_{kibc}  - g_{ka} \Theta_{ijbc} -g_{ib} \Theta_{jkca} \\
- g_{jb} \Theta_{kica} - g_{kb} \Theta_{ijca} -g_{ic} \Theta_{jkab} - g_{jc} \Theta_{kiab} - g_{kc} \Theta_{ijab}
\end{multline}
%\end{eqnarray}

\bibliographystyle{utphys}

\bibliography{algebraextensions}

\providecommand{\href}[2]{#2}\begingroup\raggedright\endgroup

\end{document}